\documentclass[a4paper,fleqn,usenatbib]{mnras}

\usepackage{newtxtext,newtxmath}

\usepackage[T1]{fontenc}
\usepackage{ae,aecompl}

\usepackage{graphicx}	
\usepackage{amsmath}	
\usepackage{amssymb}	
\usepackage{longtable}
\usepackage{lscape}
\usepackage{multirow}
\usepackage{caption}
\usepackage[usenames,dvipsnames]{xcolor}

\newcommand{\Teff}{\mbox{$T_{\mathrm{eff}}$}}
\newcommand{\kms}{\mbox{$\mathrm{km\,s^{-1}}$}}

\newcommand{\Msun}{\mbox{$\mathrm{M}_{\odot}$}}

\newcommand{\masy}{mas\,yr$^{-1}$}
\newcommand{\cpc}{pc$^3$}
\newcommand{\pcpc}{pc$^{-3}$}
\newcommand{\dd}{\mbox{$\mathrm{d}$}} 
\newcommand{\Gabs}{\mbox{$G_{\rm abs}$}}
\newcommand{\bp}{\mbox{$G_{\rm BP}$}}
\newcommand{\rp}{\mbox{$G_{\rm RP}$}}
\newcommand{\completeness}{\mbox{$96.0^{+1.3}_{-1.6}$\,percent}}
\newcommand{\WDdensity}{\mbox{$(4.49\pm0.38)\times10^{-3}$\,\pcpc}}
\newcommand{\gaia}{{\it Gaia}}

\title[The \gaia\ 20\,pc white dwarf sample]{The {\em Gaia} 20\,pc white dwarf sample}

\author[M. A. Hollands et al.]{M.~A.~Hollands,$^{1}$\thanks{E-mail: M.Hollands.1@warwick.ac.uk (MH)}
P.-E.~Tremblay,$^{1}$
B.~T.~G\"ansicke,$^{1}$
N.~P.~Gentile-Fusillo,$^{1}$
\newauthor
S. Toonen$^{2}$
\\
$^{1}$ Department of Physics, University of Warwick, Coventry CV4 7AL, UK \\
$^{2}$ Anton Pannekoek Institute for Astronomy, University of Amsterdam, 1090 GE Amsterdam, The Netherlands
}

\date{Accepted XXX. Received YYY; in original form ZZZ}

\pubyear{2018}

\begin{document}
\label{firstpage}
\pagerange{\pageref{firstpage}--\pageref{lastpage}}
\maketitle

\begin{abstract}
Using \gaia\ DR2 data, we present an up-to-date sample of white
dwarfs within 20\,pc of the Sun. 
In total we identified 139 systems in \gaia\ DR2,
nine of which are new detections, with the closest of these
located at a distance of $13.05$\,pc. 
We estimated atmospheric parameters for all stellar remnants based on the \gaia\ parallaxes
and photometry. The high precision and completeness of the \gaia\
astrometry allowed us to search for wide binary companions. We re-identified
all known binaries where both components have accurate DR2 astrometry,
and established the binarity of one of the nine newly identified white dwarfs.
No new companions were found to previously known 20\,pc white dwarfs.
Finally, we estimated the local white dwarf space-density 
to be \WDdensity, having given careful consideration to the distance-dependent
\gaia\ completeness, which misses known objects at short distances, but is close to
complete for white dwarfs near 20\,pc.
\end{abstract}




\section{Introduction}
White dwarfs are the progeny of stars with initial masses $\la8-10$\,\Msun\
\citep{garcia-berroetal97-1, ibenetal97-1, smarttetal09-1}. The strong
correlation between stellar mass and luminosity implies rapidly decreasing
main-sequence life times for stars more massive than the Sun
\citep{massey+meyer01-1}, and consequently, the vast majority of stars with
$M\ga1.5$\,\Msun\ ever formed in the Galaxy have already become white dwarfs.
Moreover, due to the steep slope of the initial mass function, most of the
present-day white dwarfs descended from $\simeq1.2$--$2.5$\,\Msun\ stars.
Consequently, the solar neighbourhood is strongly dominated ($\simeq75$\,percent) by
low-mass M and K-type stars \citep[see, e.g.,][]{finchetal14-1,henryetal18-1}. 
The remaining $\simeq25$\,percent are nearly equally
split between white dwarfs and main-sequence G, F and A-type
stars.\footnote{http://www.recons.org/census.posted.htm} Therefore, a
volume-limited sample of the stellar population provides constraints on fundamental stellar evolution owing to the rich variety of spectral types \citep{giammicheleetal12-1,limogesetal15-1,fuhrmannetal17-1}, the local star formation history \citep{wingetetal87-1,tremblayetal14-1}, the initial mass function \citep{tremblayetal16-1}, chemical evolution \citep{fuhrmannetal17-2}, and the kinematic properties of the Galactic disk or the halo \citep{sionetal14-1}. Moreover, as about half of all stars are members of binaries
or higher hierarchical multiples, such a sample also yields insight into the
initial distributions of binary mass ratios and orbital separations
\citep{duquennoy+mayor91-1, raghavenetal10-1, duchene+kraus13-1, tokovinin14-1, moe+distefano17-1, fuhrmannetal17-3, toonenetal17-1}.
The local sample is also a benchmark for the properties and frequencies of 
exoplanetary systems around stars and white dwarfs \citep{koesteretal14-1,dittmannetal17-1}.

Because of the intrinsic faintness of both white dwarfs and low-mass stars, our
knowledge of the complete stellar population is currently limited to
$\simeq10$\,pc. For white dwarfs, \citet{holbergetal02-1, holbergetal08-1,
holbergetal16-1} compiled the historic local population from the
literature, and in parallel, \citet{subasavageetal07-1, subasavageetal08-1}, \citet{sayresetal12-1} and \citet{subasavageetal17-1} pursued a dedicated search for
nearby white dwarfs, largely based on using reduced proper motions as a proxy
for distance. Both groups concluded that the stellar remnant sample within 13\,pc 
can be considered as being essentially complete.  
\citet{holbergetal08-1,holbergetal16-1} accordingly estimated the local space-density of white dwarfs at $4.8 \pm 0.5 \times 10^{-3}$\,\pcpc, corresponding to a mass of $3.3 \pm 0.3 \times 10^{-3}$\,\Msun\,\pcpc\,given the mean properties of local remnants \citep{giammicheleetal12-1}. This density is expected to hold for the 20\,pc sample since the latter volume is located well within the estimated vertical scale height of the disk even for the youngest local populations \citep{wegg+phinney12-1,buckner+froebrich14-1,joshietal16-1}. Based on that approximation, the pre-\gaia\ 20\,pc white dwarf sample is expected to be about 82--86\,percent complete \citep{tremblayetal14-1,holbergetal16-1}. In comparison, a slightly larger space-density of $5.5 \pm 0.1 \times 10^{-3}$\,\pcpc\,was estimated using the Sloan Digital Sky Survey sample corrected for completeness \citep{munnetal17-1}.
 Besides this small discrepancy, there are independent suggestions that the 20\,pc sample may be somewhat less complete than currently assumed. While a small number of single, but so far unconfirmed, nearby white dwarf candidates have been reported \citep[e.g.][]{reyleetal06-1}, the identification of white dwarf companions to main-sequence stars is a challenging problem \citep{ferrario12-1}. \citet{holbergetal13-1} discussed the statistics of spatially resolved white dwarf plus F/G/K-type stars, and commented on the fact that three out of five white dwarfs with $D<5$\,pc are within such binaries. However, even M-type stars can outshine cool white dwarfs, with a number of reported candidate binaries within 20\,pc \citep{delfosseetal99-1,maceetal18-1}.

The extreme astrometric precision of the \gaia\ Data Release 2 (DR2) will unambiguously allow the confirmation or refutation of all published single white dwarf candidate members with $D\le20$\,pc, and to potentially identify previously unknown nearby degenerate stars.  An accurate distance will also be very powerful in corroborating unresolved white dwarf companions to main-sequence stars via the detection of blue/ultraviolet excess flux. \gaia\ DR2 includes 5D phase-space parameters and colours in the $G$, \bp, and \rp\ \gaia\ passbands for a large fraction of sources brighter than $G\simeq21$ magnitude \citep{gaiaDR2-collab-1,gaiaDR2-collab-2,gaiaDR2-collab-3} with a precision that is already of the order of that expected by the end of the mission.
Even so, the relative number of \gaia\ sources with 5- vs. only 2-parameter astrometry begins to fall rapidly for sources fainter than $G\simeq 19$\,mag \citep[Fig.~2]{gaiaDR2-collab-1},
which is potentially problematic for the very coolest and thus faintest white dwarfs,
which can have absolute magnitudes exceeding $\Gabs=17$\,mag \citep[e.g. J1251$+$4404,][]{gianninasetal15-1}.
In this work we therefore focus on the white dwarf population within 20\,pc, 
which ensures we are not biased against even the faintest degenerates.
We firstly identify new white dwarfs from the \gaia\ HR diagram (Section~\ref{sec:sample}) and look at the kinematics (Section~\ref{sec:kin}) and binarity of the population (Section~\ref{sec:multiples}). We then quantify the \gaia\ selection function and resulting space-density for local white dwarfs, a sample that covers the full sky almost isotropically as well as a wide range of magnitudes and proper motions (Section~\ref{sec:spacedensity}).
We present our conclusions in Section~\ref{sec:summary}. 

\section{The \emph{Gaia} 20\,pc white dwarf sample}
\label{sec:sample}

Reddening does not pose a problem to target selection within 20\,pc. Therefore a relatively simple query of the {\it
Gaia} database is sufficient to identify all white dwarfs with full
five-parameter astrometric solutions. With the goal to inspect, and define
the local white dwarf cooling sequence in the Hertzsprung-Russell diagram (HRD), 
we began by searching the \gaia\ DR2 source catalogue for all objects 
with parallaxes greater than 40\,mas, (i.e. $D\le25$\,pc),\footnotemark\ resulting in 9284 objects. 

\footnotetext{
For the general case, calculating distances from \gaia\ parallaxes is not trivial
\citep{bailer-jones15-1, bailer-jonesetal18-1}.
However, for white dwarfs within 20\,pc, the fractional parallax uncertainties are
sufficiently small that distances and their uncertainties can be accurately 
calculated according to $D = 1/\varpi$, $\sigma_D = \sigma_\varpi/\varpi^2$.
}

\begin{figure*}
\includegraphics[width=1.8\columnwidth]{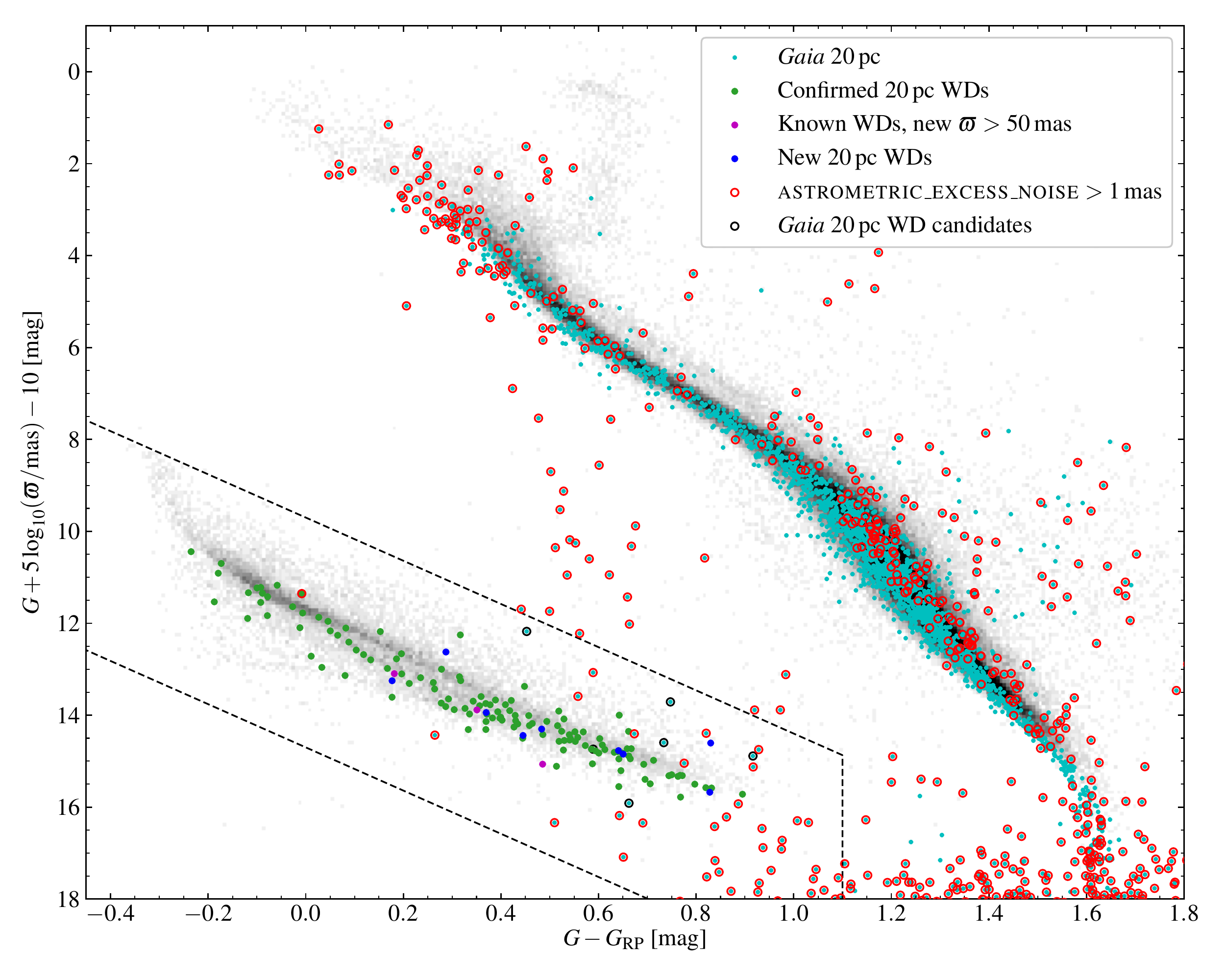}
\caption{\label{fig:hrd} Hertzsprung-Russell diagram of the stellar population within 20\,pc detected by 
\textit{Gaia}, non-degenerate stars are plotted as small turquoise symbols.
White dwarfs previously known to be within 20\,pc are shown in green, 
known white dwarfs that the \textit{Gaia} parallax places at $D<20$\,pc 
in magenta, and newly discovered degenerate stars in blue. The dashed lines show
the colour cut used to select white dwarf candidates. \textit{Gaia} sources with
$\textsc{astrometric\_excess\_noise} > 1$\,mas are indicated by red circles --
only one known white dwarf (Sirius\,B) is affected by this.
The six  rejected white dwarf candidates are shown by black circles,
though one of these is partially obscured by known white dwarfs.
For guidance, clean main-sequence and white dwarf sequences selected
from \textit{Gaia} DR2 are shown in grey.
}
\end{figure*}

We then cross-matched an up-to-date list of 193 confirmed white dwarfs, based on 
\citet{toonenetal17-1} with revisions from the recent literature 
\citep{holbergetal16-1,finchetal16-1,kirkpatricketal16-1,tremblayetal17-1,
subasavageetal17-1,finchetal18-1}, 
with published distance estimates that placed them, within uncertainties, within 20\,pc. 
Applying a \textit{Gaia} parallax cut,\footnotemark
\begin{equation}
\varpi + 3\sigma_\varpi > 50~,
\label{Eq1}
\end{equation}
i.e. including white dwarfs that are within their uncertainties at $D<20$\,pc,
we recovered 125 previously known 20\,pc white dwarfs, indicated by their WD numbers 
in Table~\ref{tab:wd20_gaia} and shown as green dots in Fig.~\ref{fig:hrd}.  
Additionally, 57 white dwarfs of the 193 were found to have distances beyond 20\,pc
(Table~\ref{tab:out20}) and 11 stellar remnants were either absent in \gaia\ DR2, or did
not have full five-parameter astrometry (discussed further in Section~\ref{sec:missing}).

\footnotetext{We decided to adopt the published  parallax zero-point, the 0.03\,mas offset
discussed by \citet{gaiaDR2-collab-2} is of the order of the $1\,\sigma$ uncertainty
on $\varpi$  for the bulk of the white dwarf sample.}

These 125 stars were used to define a generous colour cut to select a sub-set of
the \textit{Gaia} 20\,pc sample that we would scrutinize for so far unknown white dwarfs,  
\begin{eqnarray}
\Gabs & > & 9.7 + 4.7 \times (G-\rp)~, \\
\Gabs & < & 14.7 + 4.7 \times (G-\rp)~, \\
G-\rp < 1.1~, \\
\varpi + 3\sigma_\varpi & > & 50~. \label{Eq5}
\end{eqnarray}
We use $G-\rp$ as colour information instead of $\bp-\rp$, as very low mass main-sequence
stars have poorly defined \bp\ magnitudes, and consequently scatter into the white
dwarf locus in the HRD. 
The HRD selected with the above cuts contains a large number of objects between the faint ends 
of the main-sequence and the white dwarf cooling sequence, at suspiciously 
close distances ($<5$\,pc). Inspection of the spatial distribution revealed that the 
majority of these sources are in the Galactic plane. Given that for the 20\,pc sample 
the spatial distribution should be isotropic demonstrates that these sources correspond 
to spurious detections. Applying a simple cut on the astrometric noise 
($\textsc{astrometric\_excess\_noise} < 1$\,mas) similarly to what has been done in 
\citet{gaiaDR2-collab-5} removes the majority of these problematic data (flagged with red outlines
in Fig.~\ref{fig:hrd}).  
This cut removes, however, one known white dwarf, Sirius~B (WD\,0642$-$166). Given that Sirius~B 
is the companion to the brightest known star in the sky apart from the Sun, we believe it is adequate
to treat this system differently and add it to our sample.\footnotemark\
We have not applied any of the other quality cuts discussed in \citet{gaiaDR2-collab-5}. 
In particular, the suggested cut on \textsc{visibility\_periods\_used} would have removed five
known white dwarfs for which the \gaia\ data agree with ground-based measurements and one new degenerate star.

\footnotetext{Similarly, the Procyon system is absent from \gaia\ DR2 possibly from
saturation of the primary, and 40\,Eri\,B is also missing though the A and C
components are present in DR2 (see Section~\ref{sec:missing}).}

Removing the 125 known white dwarfs left 20 objects that we inspected 
individually. Three of them turned out to be known white dwarfs that were not previously considered 
to be within 20\,pc (WD\,0959$+$149: $25.05\pm0.69$\,pc, \citealt{holbergetal16-1};
WD\,1316$-$215: $31.6\pm5.3$\,pc, \citealt{subasavageetal07-1};
WD\,1350$-$090:  $25.3\pm1.0$\,pc, \citealt{giammicheleetal12-1}).
Two more of the 20 were very recent white dwarfs identifications that were missed from our
initial list of 193 objects. One (\gaia\ DR2 5867776696271127424) was found as a probable white dwarf by
\citet{smithetal18-1} in the VVV survey, owing to its large reduced-proper motion.
The \gaia\ DR2 parallax secures the degenerate nature of this star.
The other (\gaia\ DR2 2202703050401536000) was recently identified by \citet{scholzetal18-1} as a
companion to TYC\,3980-1081-1, located at a distance of 8.46\,pc.
For the remaining 15 objects, we 
analysed archival photometry and imaging, in particular from Pan-STARRS and 2MASS.
The nature of these objects is discussed in detail in Section\,\ref{sec:newWDs}, where we conclude 
that 9 of them are indeed white dwarfs (blue dots in Fig.~\ref{fig:hrd}), whereas 
the final 6 are likely contaminants scattered into the white dwarf sequence in the HRD due to 
poor photometry and/or astrometry. 

\subsection{Atmospheric parameters}

\label{sec:atmfit}

In order to fully understand the \gaia\ DR2 data set for known white dwarfs and to
prepare for the identification of new local remnants in Section\,\ref{sec:newWDs},
we estimated the atmospheric parameters of all sources that are included in our cuts
given by Equations~(\ref{Eq1}--\ref{Eq5}). We only employed the measured fluxes in
the $G$, \bp\ and \rp\ passbands as well as the parallax as input parameters of our fitting
procedure. We have used pure-hydrogen \citep{tremblayetal11-1}, pure-helium \citep{bergeronetal11-1},
and mixed model atmospheres \citep{tremblayetal14-2} to calculate the
monochromatic surface Eddington flux $H_\lambda$ as a function of effective
temperature (\Teff) and surface gravity ($\log g$). The mass-radius relations of
\citet{fontaineetal01-1} for thick hydrogen layers ($M_{\rm H}/M_{\rm WD}$ =
10$^{-4}$; pure-H atmospheres) and thin hydrogen layers ($M_{\rm H}/M_{\rm WD}$ =
10$^{-10}$; pure-He and mixed atmospheres)
were employed to compute the stellar flux of the white dwarf as a point source,
which was then integrated through the revised \gaia\ DR2
passbands $S(\lambda)$ of \citet{gaiaDR2-collab-3} in units of quantum
efficiency as
\begin{equation}
  F^m_\lambda = \frac{\int 4\pi R^2 H_\lambda(\Teff,\log g) S^m(\lambda) \lambda \,\dd\lambda}
  {\int S^m(\lambda)\lambda \,\dd\lambda} ~,
  \label{fit1}
\end{equation}
where $m$ represents a given \gaia\ filter ($G$, \bp, \rp). The observed quantity is 
\begin{equation}
  f^{\rm m}_\lambda = \varpi^2 ~ F^{m}_\lambda~,
  \label{eqZeroPoint3}
\end{equation}
{\noindent}where $f^{\rm m}_\lambda$ is the \gaia\ flux in units of erg\,cm$^{-2}$\,s$^{-1}$.
To determine the \Teff\ and $\log g$, we minimized the $\chi^2$ between
the absolute \gaia\ photometry and absolute synthetic magnitudes calculated
from the above models, employing the non-linear least-squares
Levenberg-Marquardt algorithm \citep{pressetal92-1}.
The stellar radius, $R$, in Equation~(\ref{fit1}) is fully defined by
the atmospheric parameters and our adopted mass-radius relation. The
uncertainties were directly obtained from the covariance matrix of the fit although these are underestimates, since we did not include
the unknown systematic errors of the spectral models, \gaia\ passbands,
or mass-radius relations in the $\chi^2$.

The currently known stellar remnants within 20\,pc have a wide variety of
atmospheric compositions and spectral types, summarised in Table~\ref{tab:wd20_gaia}.
While the majority of these white dwarfs are thought to have pure hydrogen or pure helium
atmospheric compositions,
at least 39\,percent have a magnetic field, or traces 
of carbon and other metals. As a consequence, a complete model atmosphere
analysis using all available spectroscopy as well as new \gaia\ data would
be necessary to provide a physically meaningful 
update of the atmospheric parameters of
previously known white dwarfs. The \gaia\ solutions presented in
Table~\ref{tab:wd20_gaia} should be taken with caution and serve as a reference to understand atmospheric parameters derived from \gaia\  and the future characterisations of larger \gaia\ samples.

\begin{figure}
\includegraphics[width=\columnwidth]{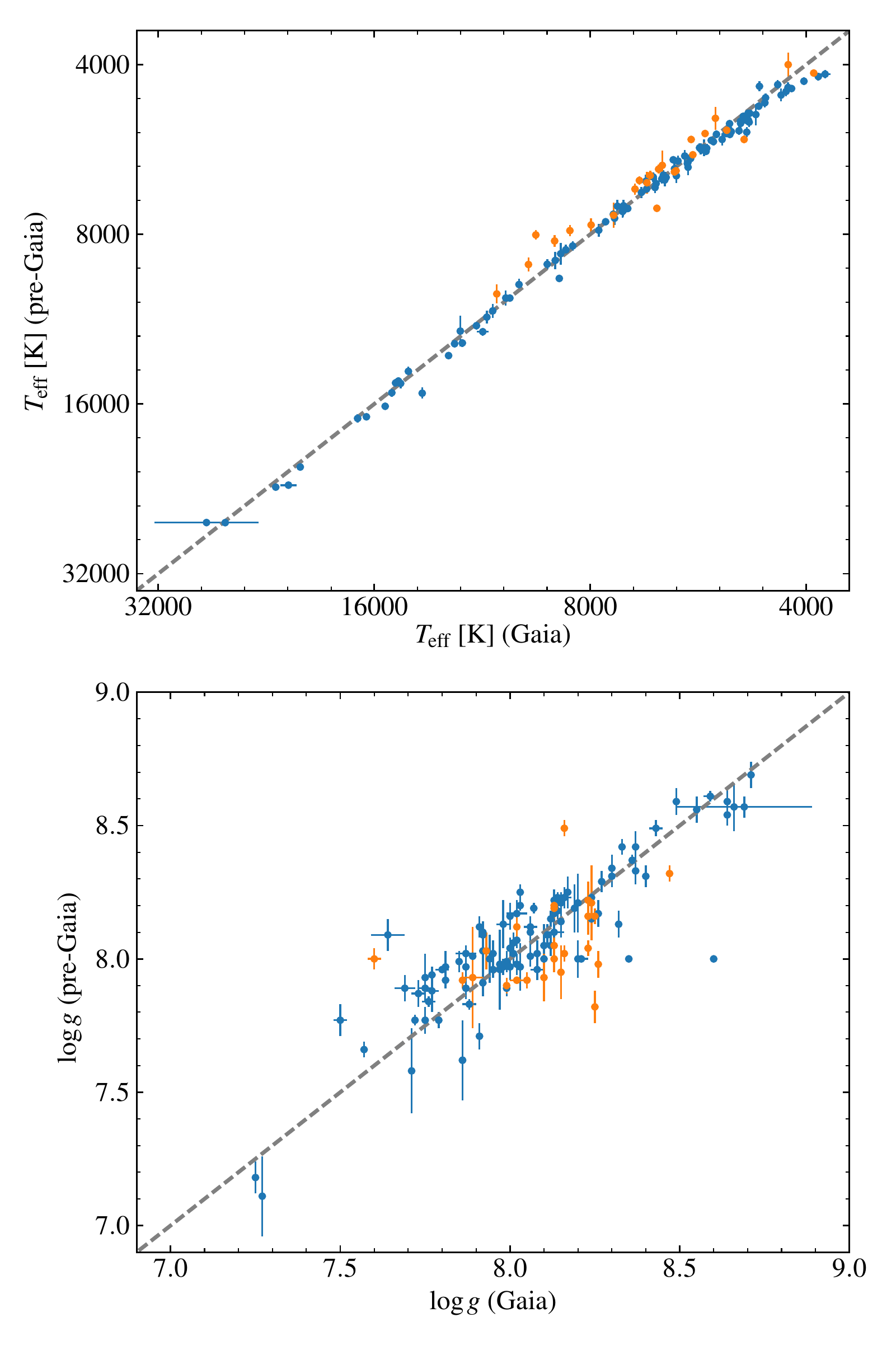}
\caption{Comparison of published atmospheric parameters and those derived here based on the \gaia\ parallaxes and photometry. DQ and DZ stars are shown in orange with all other objects in blue.\label{parameters1}}
\end{figure}

\begin{figure}
\includegraphics[width=\columnwidth]{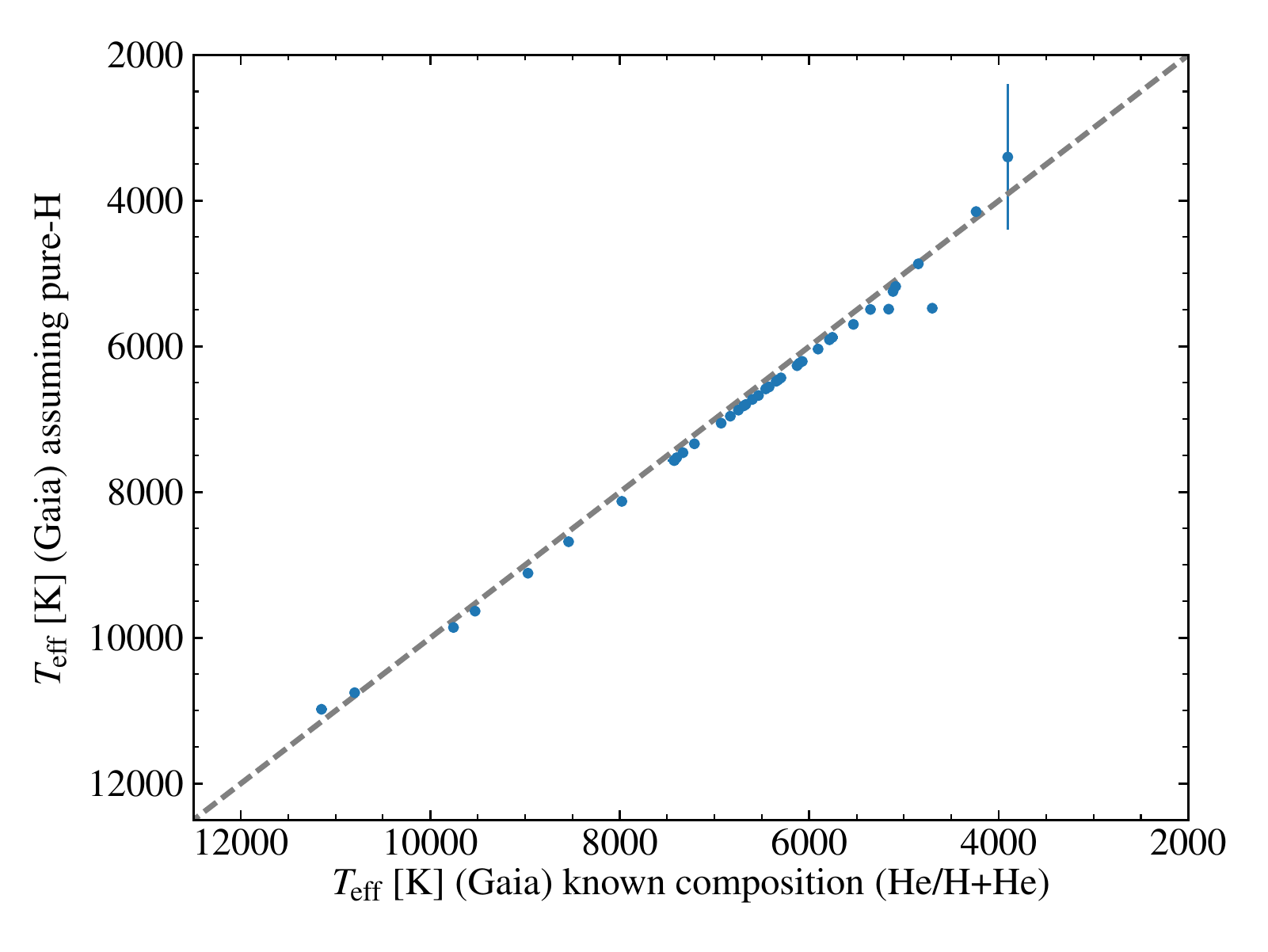}
\caption{Comparison of the fits to the \gaia\ data assuming a pure-H atmosphere for those objects in Table~\ref{tab:wd20_gaia} that are assigned a pure-He or mixed He/H atmosphere. \label{parameters2}}
\end{figure}

In Fig.~\ref{parameters1} we highlight the comparison of our \gaia\ DR2 parameters
to previously established values from the literature 
\citep{gianninasetal11-1,farihietal11-3,giammicheleetal12-1,limogesetal15-1,subasavageetal17-1}.
As much as possible we adopted literature parameters based on
photometric rather than spectroscopic estimates for objects with $\Teff < 10\,000$\,K
allowing for a more direct comparison with \gaia. In the case of 1D spectroscopic solutions
we corrected for 3D effects \citep{tremblayetal13-1}. In all cases \gaia\ DR2 alone can not
constrain the atmospheric composition and we adopt in Table~\ref{tab:wd20_gaia} the same
composition as these earlier analyses. For white dwarfs of unknown spectral types,
we assumed a pure-hydrogen composition. Fig.~\ref{parameters1} suggests a very good
agreement between established and \gaia\ DR2 parameters, even for DQ and DZ white dwarfs
(orange), which differ significantly from the pure-helium approximation. This is likely a 
consequence of the extremely broad \gaia\ passbands and the fact that the sensitivity of atmospheric
parameters to \gaia\ colours decreases significantly with temperature. Fig.~\ref{parameters2} 
compares our adopted \gaia\ DR2 parameters with an analysis that assumed a pure-hydrogen
composition for all objects with helium dominated atmospheres. The effect is fairly mild owing to the
broad \gaia\ passbands, and it suggests that using the pure-hydrogen approximation for newly discovered
\gaia\ white dwarfs provides a fairly reasonable estimate of their atmospheric parameters
as long as one is cautious about possible outliers when deriving astrophysical relations.

\begin{figure}
\includegraphics[width=\columnwidth]{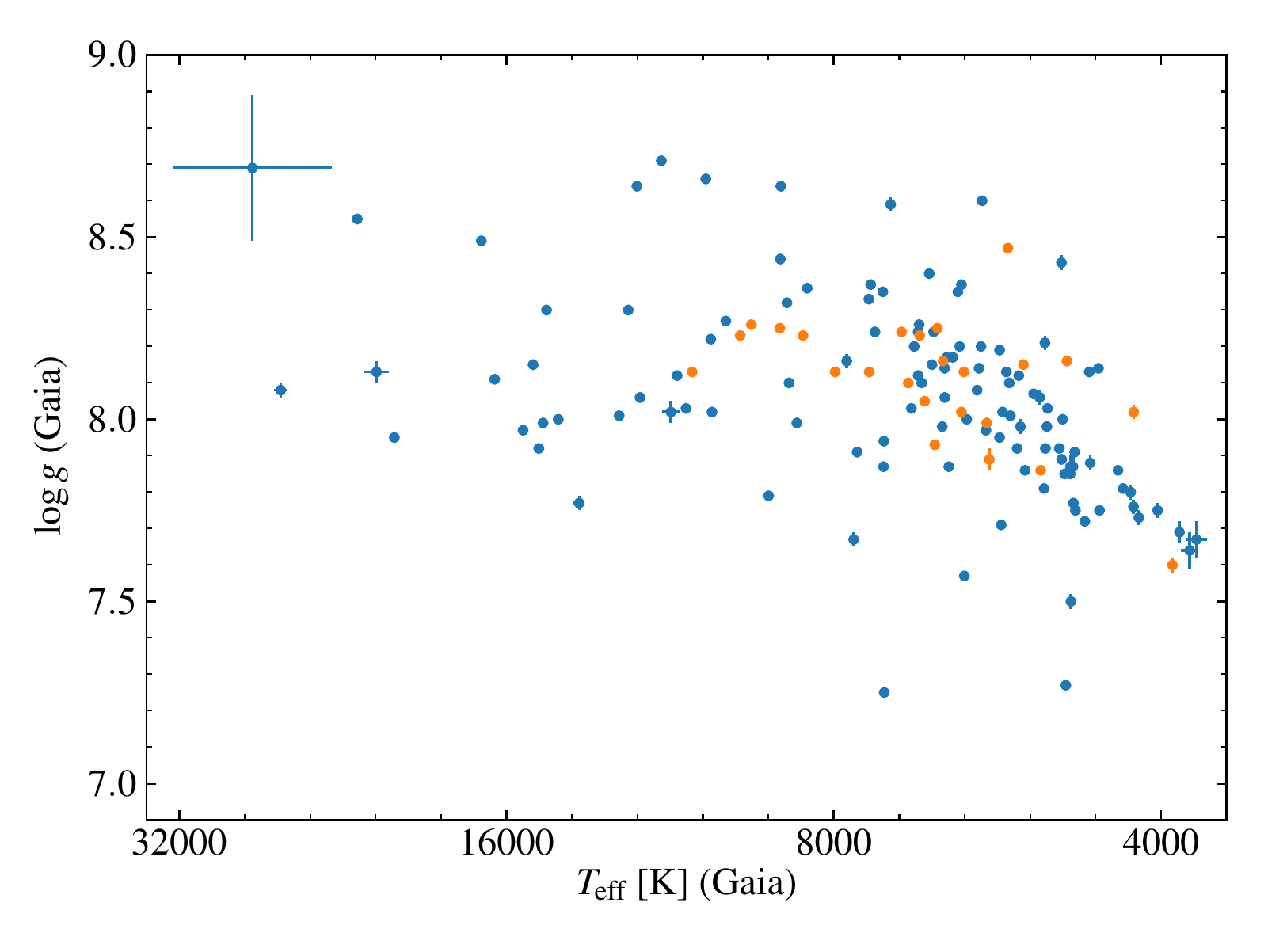}
\caption{$\log g$ - \Teff\ distribution of photometric \gaia\ DR2 parameters. The colours of the symbols are the same as in Fig.~\ref{parameters1}.\label{parameters3}}
\end{figure}

The precision of the \gaia\ DR2 data is extremely high resulting in similarly small error bars on the atmospheric parameters in Table~\ref{tab:wd20_gaia} and Figs.~\ref{parameters1}--\ref{parameters3}. It is clear that these error bars are only statistical in nature, and should be taken with some caution, especially considering our approximation on the atmospheric composition. Furthermore, it is unlikely that the accuracy of the physics in the model atmospheres reach the \gaia\ 20\,pc sample precision. A much more conservative error estimate for the atmospheric parameters of individual objects can be made by comparing the average deviation between \gaia\ and published parameters. We find deviations of 3.1 percent in \Teff\ and 0.10\,dex in $\log g$. This likely overestimates the errors since the \gaia\ data set is more precise and homogeneous than earlier measurements.

The \gaia\ DR2 surface gravity distribution of the 20\,pc sample is presented in Fig.~\ref{parameters3}. It agrees with a sharply peaked distribution centred on the canonical value of $\log g = 8.0$. However, the coolest objects in the sample, adopting the atmospheric composition of published photometric analyses, show a trend towards lower surface gravities. We currently do not know the reason for this behaviour that was also observed by \citet{gaiaDR2-collab-5}. In this regime, the effects of collision-induced-absorption and the red-wing of the Ly$\alpha$ line become important for the optical colours \citep{borysowetal01-1,kowalski+saumon06-1,blouinetal17-1}.

\subsection{New white dwarfs}
\label{sec:newWDs}

The colour cuts of Equations~(\ref{Eq1}--\ref{Eq5}) and the quality cut on astrometric noise left us with 15 new white warfs candidates within 20\,pc. The \gaia\ photometric fits performed in Section~\ref{sec:atmfit} can help to identify the nature of these objects but additional information was needed to separate genuine new white dwarfs from spurious entries in the \gaia\ database. To this aim, we have gathered additional photometry by matching 13 objects with 2MASS \citep{skrutskieetal06-1} and 8 with Pan-STARRS \citep{chambersetal16-1}. We have integrated the same model grids as those used in Section~\ref{sec:atmfit} to calculate synthetic fluxes using the appropriate filters \citep{cohenetal03-1,tonryetal12-1}. In Appendix~\ref{app_A} we show all individual fits for 8 objects with both Pan-STARRS and 2MASS, 5 sources with \gaia\ and 2MASS, and 2 stars with \gaia\ data only. For 9 objects we conclude that all available data is consistent with a previously unknown white dwarf, these stars are flagged accordingly in Table~\ref{tab:wd20_gaia}. The table also includes their estimated atmospheric parameters derived from \gaia\ data alone, consistent with the sample of known white dwarfs. A more detailed assessment of their properties will have to await follow-up spectroscopy. The following stars are worth a special mention:

{\bf Gaia DR2 2486388560866377856} 
is a high-proper motion star (LP~649$-$66), 
and has a common proper motion M-dwarf companion (LP~649$-$67, see Section~\ref{sec:widebin}).  
Neither Pan-STARRS or 2MASS data  are available, but the low luminosity and \gaia\ 
colours of this source confirm the degenerate nature.

{\bf Gaia DR2 4970215770740383616} is an outlier in terms of its 
\textsc{phot\_bp\_rp\_excess\_factor} value of 1.661, indicating that the modelling of the background 
flux is poor. This star is a known high-proper motion object \citep{luyten+hughes80-1, 
plataisetal98-1}, which is well-resolved in archival imaging, but moved to within $\simeq1.1$\,arcsec 
of a star of nearly identical optical brightness during the epoch of the \textit{Gaia} 
observations (DR2), which probably explains the poor astrometry. However, considering the 
high proper motion and the fact that the \gaia\ $G$ and $JHK$ magnitudes agree with a degenerate
star of $\Teff = 5270$\,K and $\log g = 8.0$, we conclude that it is likely a new white dwarf. 
Ground-based follow-up spectroscopy will be challenging for the next few years due to its close
proximity to the other star.

{\bf Gaia DR2 5224999346778496128} is a faint ($G$ = 17.18 mag), high-proper motion source, that is neither in 2MASS or Pan-STARRS. It has no \gaia\ flags that would suggest lower quality data hence we consider this object as a very cool white dwarf. 

We are left with 6 \gaia\ sources that are unlikely to be white dwarfs which are summarised
in Table~\ref{tab:rejected}. Five of them have $JHK$ flux excesses suggesting a main-sequence star.
The average \textsc{astrometric\_excess\_noise} parameter for these objects is 0.56\,mas,
while the average for the new and confirmed 20\,pc white dwarf sample is much lower at 0.07\,mas.
Furthermore, all but one of these objects have $\textsc{visibility\_periods\_used}<8$
and hence would not have made the cuts proposed in \citet{gaiaDR2-collab-5}.
A deeper discussion is required for the rejected object Gaia DR2 6206558253342895488.
Inspection of its photometric fit (Fig.~\ref{fitsA2}) shows mild disagreement between the \gaia\ and
2MASS photometry. This is most likely contamination from a B-type star situated 16\,arcsec away,
seeing as the 2MASS contamination flags are raised. Visually, the data resembles
WD\,2307$+$548, whose 2MASS photometry is similarly contaminated by its companion.
While, on this basis, claiming \gaia\ DR2 6206558253342895488 to be a new white dwarf is tempting,
the accuracy of its astrometry is questionable. The astrometric excess noise is a modest 0.62\,mas,
however the proper motion is the most suspicious aspect.
The \gaia\ data give $(\mu_\mathrm{Ra}, \mu_\mathrm{Dec}) = (+71.19\pm0.82, -54.16\pm0.56)$\,\masy,
but is found to be $(-36.5\pm9.0, -25.3\pm9.0)$\,\masy\ and $(-74.9\pm6.9, -28.0\pm6.6)$\,\masy\
in the NOMAD \citep{zachariasetal04-2} and SPM \citep{girardetal11-1} catalogues, respectively.
While these two measurements are only marginally consistent with each other, they are both in
clear disagreement with the \gaia\ proper motion. Therefore it reasonable to assume that the
\gaia\ parallax of this star is equally unreliable, in which case the object could be a K-type
star situated at a much greater distance.
A spectroscopic follow-up of the 6 sources identified in Table~\ref{tab:rejected}
can help to identify their nature.

\subsection{Known missing systems}
\label{sec:missing}

While \gaia\ DR2 includes the majority of known 20\,pc members, eleven white dwarfs known
(or thought to be) within 20\,pc are missing and are summarised in
Table~\ref{tab:notInDR2}.\footnote{We note in passing LHS\,1249, which was discussed by 
\citet{reyleetal06-1}  as a white dwarf candidate, based on a featureless spectrum (their Fig.~3), with
a photometrically estimated distance of 14.4\,pc. 
LHS\,1249 is catalogued as an M3-dwarf \citep{pesch+sanduleak78-1}, which 
is consistent with the \textit{Gaia} parallax and 2MASS $JHK$ photometry. We conclude
that LHS\,1249 is not a white dwarf, though the spectrum presented by \citet{reyleetal06-1}
remains puzzling.}
Of these eleven, the eight with prior parallaxes are all located within 14\,pc indicating
that non-detection primarily affects the nearest objects.
These absences appear to be caused for a variety of reasons.
For many of these, high proper motion is the probable culprit where \gaia\ DR2 is
estimated to be missing 17\,percent of sources with proper motions $>600$\,\masy\
\citep{gaiaDR2-collab-1}, although this is described as having a greater impact
on bright stars.
Binary companions can also lead to the non-detections of white dwarfs.
In the case of Procyon\,B (WD\,0736$+$053), no sources are detected within 45\,arcsec of the primary.
While this could be attributed to saturation of Procyon\,A, we note that Sirius\,B is present
in the \gaia\ data despite its proximity to the brightest star in the sky.
For the 40\,Eri system, the white dwarf component (40\,Eri\,B) is missing although
the A and C components both have full \gaia\ astrometry.
For the WD\,0727$+$482A/B system, the non-detection of both white dwarfs is likely associated
with their large astrometric excess noise of $\simeq 10$\,mas, potentially caused by their
close binary orbit (their projected separation is $\simeq 7$\,au).
The missing systems are analysed further in Section~\ref{sec:spacedensity}, where they are
used to constrain the distance-dependent completeness of the \gaia\ stellar remnants within 20\,pc.

\subsection{Mass distribution}
\label{sec:mass}

The present day distribution of white dwarf masses encodes the product of
multiple astrophysical processes of scientific importance,
in particular the initial mass function, star formation history, initial-to-final mass relation (IFMR),
and binary interactions such as accretion or mergers \citep[see, e.g.,][]{liebertetal05-1,giammicheleetal12-1,tremblayetal14-1,tremblayetal16-1,toonenetal17-1}.

\begin{figure}
	\includegraphics[width=\columnwidth]{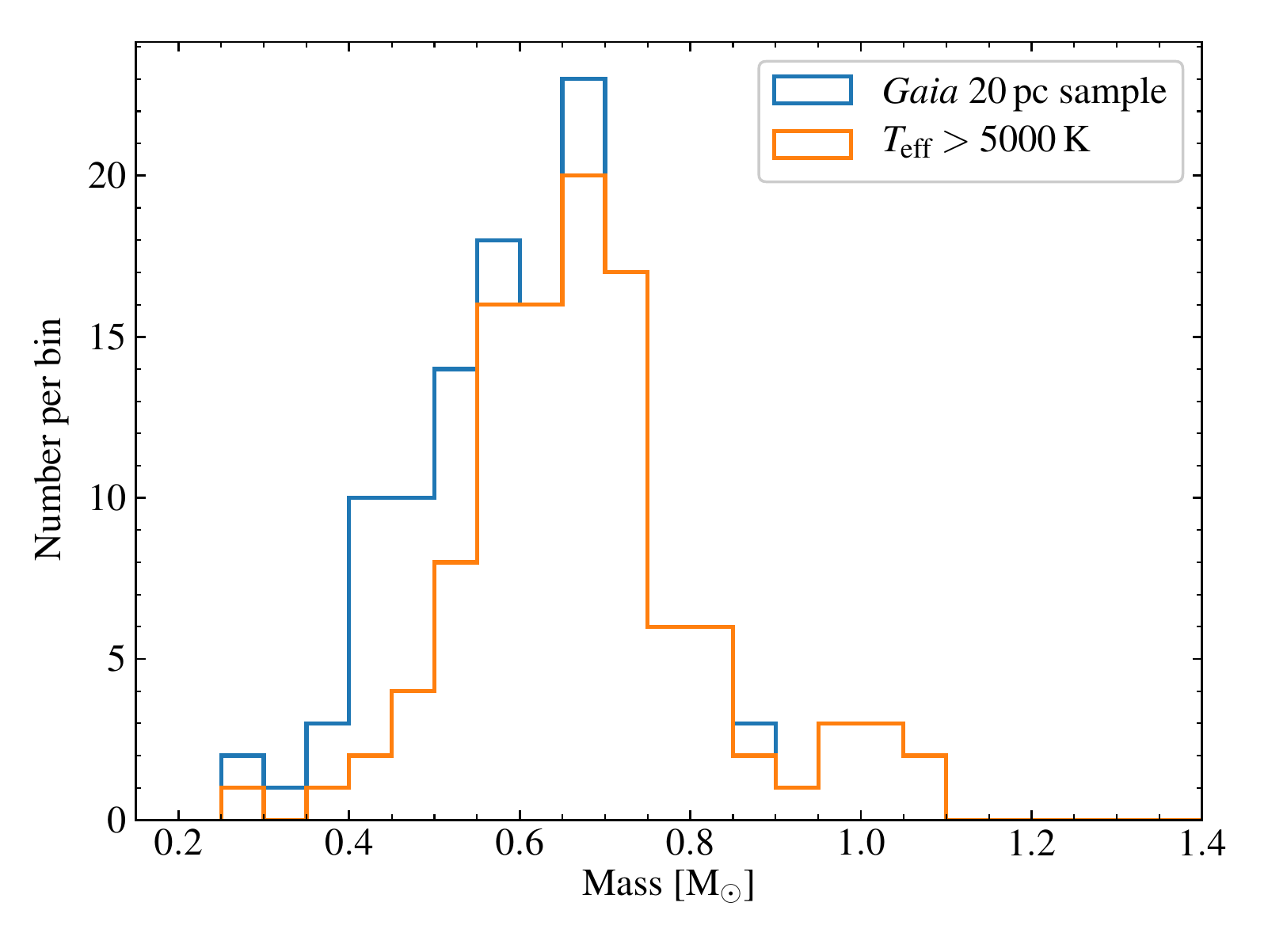}
    \caption{Mass distribution of our sample (blue). The reduced sample of objects with
    $\Teff > 5000$\,K is shown in orange.}
    \label{fig:mass}
\end{figure}

Based on the  atmospheric parameters measured from fitting the \gaia\ data, 
and our  adopted M-R relations (Section~\ref{sec:atmfit}), we determined
the mass of each white dwarf detected by \gaia\ in the 20\,pc sample. In Fig.~\ref{fig:mass} we show the mass distribution for these stellar remnants (blue). Because Fig.~\ref{parameters3} shows a systematic decrease in $\log g$ at low \Teff\
(described in Section~\ref{sec:atmfit}), we therefore present a reduced mass distribution
for objects with $\Teff > 5000$\,K (orange), which are minimally affected.

Since the release of \gaia\ DR2, \citet{kilicetal18-1} have also studied the white 
dwarf mass distribution within a distance of 100\,pc. Their much larger sample of $\simeq 14\,000$ sources showed a highly bimodal mass distribution, with 
a highest peak located at the canonical 0.6\,\Msun, plus a broader and weaker peak 
near 0.8\,\Msun, which \citealt{kilicetal18-1} attributed to mergers.

\citet{el-badryetal18-1} also analysed  white dwarfs within 100\,pc for the purpose of constraining the initial-to-final mass relation. They too identified 
a second peak in the white dwarf mass distribution located at 0.8\,\Msun,
though their analysis is limited to objects with $\Gabs < 14$.
Instead of mergers, \citet{el-badryetal18-1} attribute this to a flattening in the IFMR for 
initial masses of 3.5--5.5\,\Msun.

In contrast with the results of \citet{kilicetal18-1} and \citet{el-badryetal18-1}, Fig.~\ref{fig:mass}
does not show an excess of objects near 0.8\,\Msun, and instead appears broadly similar
to the results presented in other recent work
\citep[e.g.][]{genestetal14-1, rebassa-mansergasetal15-1, kepleretal16-1,tremblayetal16-1}.
Even considering the relatively small sample size, 
the bimodal mass distribution found by \citet{kilicetal18-1} and 
\citet{el-badryetal18-1} should be visible in Fig.~\ref{fig:mass} 
if present in the 20\,pc sample. 

\section{Kinematics}
\label{sec:kin}

The \gaia\ astrometry includes both parallaxes and proper motions. Thus the
tangential velocity, $v_\perp$, can be determined entirely from \gaia\ DR2
data.

\begin{figure}
	\includegraphics[width=\columnwidth]{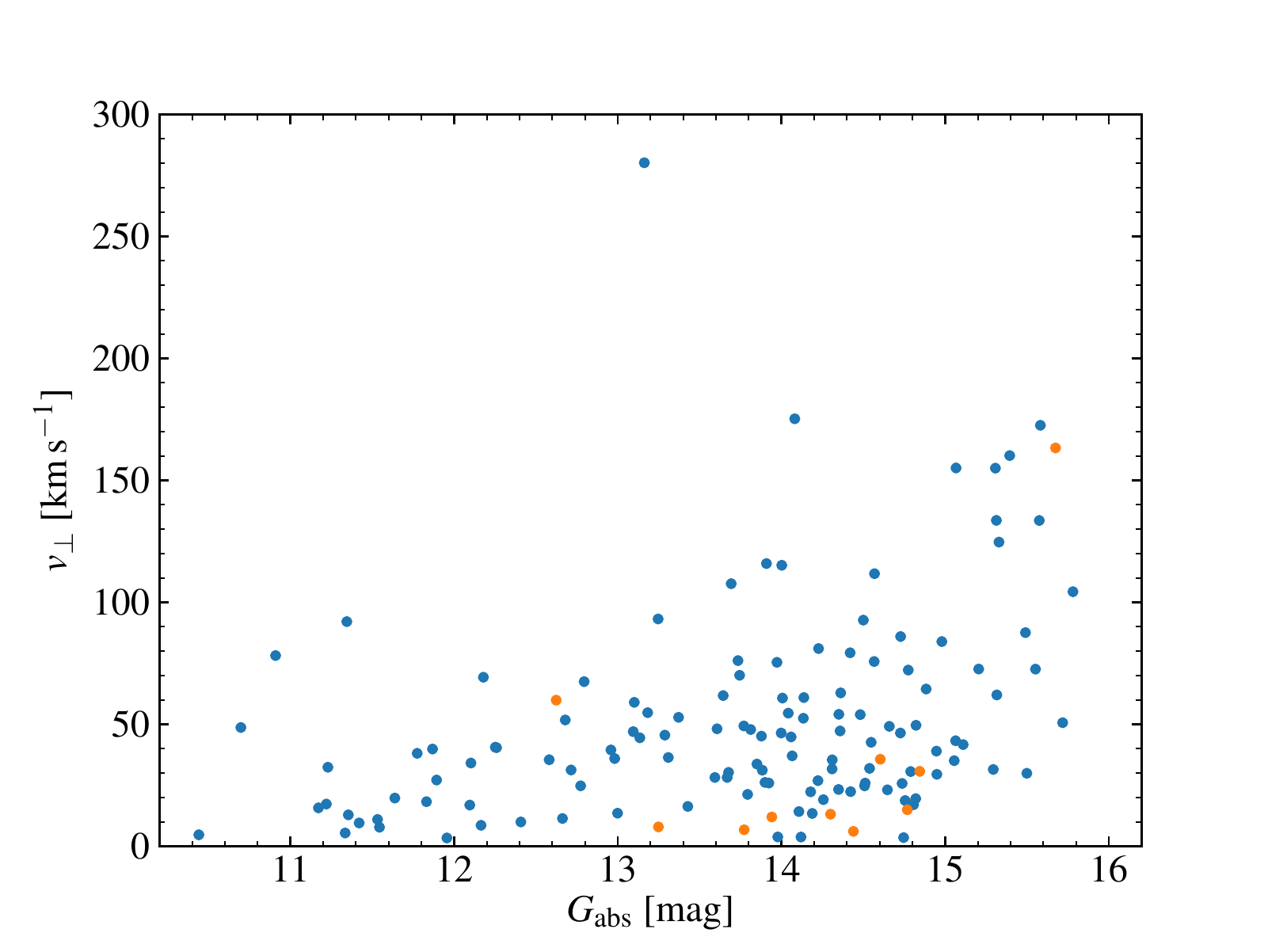}
    \caption{Tangential velocities as a function of absolute $G$ magnitude.
    The 9 new objects identified in this work are shown in orange.    }
    \label{fig:kinematics}
\end{figure}

The tangential velocity is best studied as a function of the total age of a
white dwarf including the main-sequence lifetime \citep[see, e.g., Fig. 5
of][]{tremblayetal14-1}. Using the white dwarf cooling age (or absolute
magnitude) alone, it is difficult to separate Galactic disk white dwarfs into
components that possibly have a different scale height \citep{sionetal14-1}. Nevertheless, a correlation between absolute $G$ magnitude and tangential
velocity is expected, as faint local white dwarfs with large cooling ages must have a large total age. We must be cautious, however, since young remnants are a mix of young and old stars of different masses. Fig.~\ref{fig:kinematics} shows \Gabs\ as a function of tangential velocity and confirms that correlation. All white dwarfs including new detections are consistent with disk kinematics, except possibly for WD\,1756$+$827 with $v_\perp$ = 280\,km\,s$^{-1}$. 

While most \gaia\ white dwarfs lack radial velocity measurements an estimate of the 3D velocity in terms of the U, V, and W Galactic coordinates can help to identify halo candidates \citep{chiba+beers00-1}. For WD\,1756$+$827, \citet{fuchs+jahreiss98-1} and \citet{fuhrmannetal12-1} suggest either halo membership or a high-velocity thick-disc star. WD\,1756$+$827 is a relatively young white dwarf with a cooling age of 1.4\,Gyr. However, \gaia\ parameters suggest a mass of 0.52\,\Msun, which is consistent with the value of 0.53\,\Msun\ from \citet{limogesetal15-1}, and thus a large main-sequence lifetime. 

\section{Multiple systems}
\label{sec:multiples}

With the arrival of parallaxes for an unprecedented number of stars comes the
opportunity to identify new binary/multiple systems, at both close and wide
separations. Here we outline our detections of multiple systems identified
within the 20\,pc white dwarf sample.

\subsection{Wide binaries}
\label{sec:widebin}

In the pre-\gaia\ era, a typical approach to identify resolved binaries was
to perform a cone-search around one's target, and then search for other sources
with comparable proper motion. This approach is reasonable in the absence of
universal parallaxes, but is restrictive for finding the widest binaries, particularly
for stars in the Solar neighbourhood, like those of interest here. For
example, a cone-search of a few arcmin might be chosen to avoid selecting too
large a number of sources (which could not previously be pre-filtered by parallax),
yet on-sky separations of a few degrees are possible in the most extreme cases,
where physical separations can approach 1\,pc \citep{caballero10-1}.

With the parallaxes provided by \gaia\ DR2, such limitations are relegated
to the past (at least for local stars). Instead we perform our companion search
in terms of distances and tangential-velocities.

\begin{figure}
	\includegraphics[width=\columnwidth]{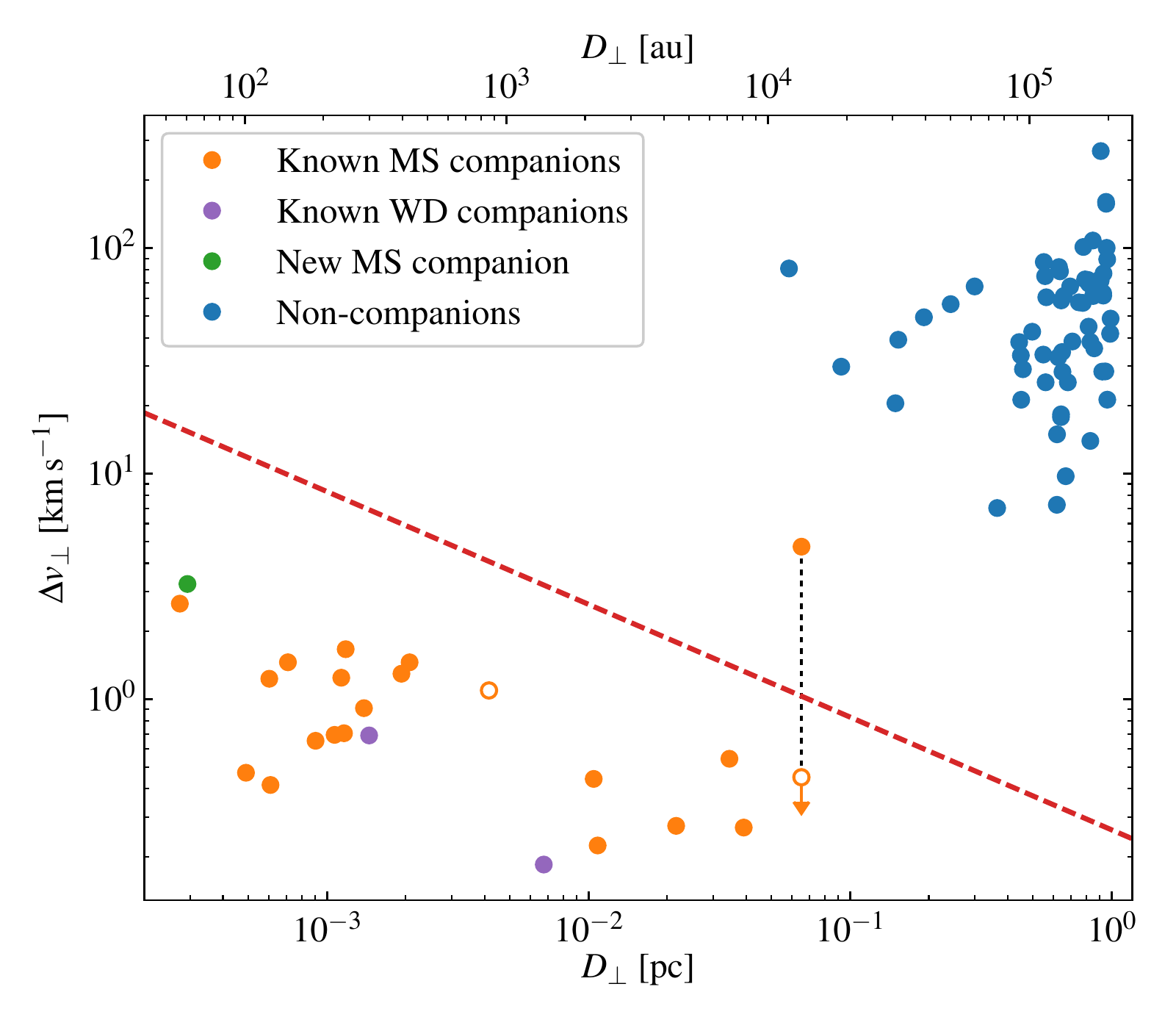}
    \caption{
    Tangential velocity differences as a function of projected separation
    for all \gaia\ sources within 1\,pc of the white dwarfs in our 20\,pc sample.
    The \gaia\ uncertainties in both $D_\perp$ and $\Delta v_\perp$ are smaller than
    the size of the points.
    For WD\,0743$-$336/171\,Pup\,A, the nominal \gaia\ value of $\Delta v_\perp$ is
    shown by the solid point, which is joined to the corrected upper limit calculated from
    PPMXL data. The hollow point at $D_\perp = 859$\,au corresponds to
    WD\,2057$-$493/WISEA\,J210104.88$-$490626.5, which is not directly recovered by
    our search due to inaccuracy of the companion parallax.
    The red dashed line indicates the maximum $\Delta v_\perp$ for a binary system containing
    two 1\,\Msun\ stars on circular orbits with a semimajor-axis of $D_\perp/2$.
    }
    \label{fig:Dvdiff}
\end{figure}

For computational efficiency, we initially limited our search to \gaia\
sources with $\varpi > 40$\,mas only.
For each white dwarf identified in Section~\ref{sec:sample}, we performed
a cone-search scaled by white dwarf distance -- in effect a cylindrical search within
a projected separation ($D_\perp$) of 1\,pc.
At 20\,pc, this corresponds to an angular-radius of about 3\,degrees.
We next made a cut on the absolute difference in radial distances ($\Delta D_\parallel$)
again of 1\,pc, but including some leeway for uncertainty, i.e.
\begin{equation}
	| \Delta D_\parallel | - 3\sigma_{\Delta D_\parallel} < 1\,\mathrm{pc}.
    \label{eq:ddparallel}
\end{equation}
Thus, for each degenerate star, we searched for companions in a cylindrical volume of
$\simeq 6.3$\,\cpc. Finally, for each white dwarf we checked the stars within
the search-volume for consistent tangential-velocities, by calculating
the velocity differences in \kms,
\begin{equation}
  \Delta v_\perp = 4.7405\sqrt{(\Delta \mu_\mathrm{Ra})^2 + (\Delta \mu_\mathrm{Dec})^2}/\varpi_\mathrm{wd}~,
\end{equation}
where $\varpi_\mathrm{wd}$ is the parallax of the white dwarf in mas, and
$\Delta\mu_\mathrm{Ra/Dec}$ are the differences in the
right-ascension/declination components of proper motion in \masy.
In Fig.~\ref{fig:Dvdiff}, we show $\Delta v_\perp$ against $D_\perp$ for all
\gaia\ sources found within the search volumes of all 20\,pc white dwarfs.\footnotemark\
Two prominent distributions are observed -- well separated stars with
large $\Delta v_\perp$ (blue points), and closer stars with more comparable $v_\perp$.
This latter group predominantly consists of known companions to 20\,pc white dwarfs,
but also a companion to one of the nine newly identified white dwarfs
(green point), which is discussed further below.
Information regarding all identified wide pairs is also presented in Table~\ref{tab:wide_bin}.

\footnotetext{For the abscissae in Fig.~\ref{fig:Dvdiff},
$D_\perp$ is preferable to either $|\Delta D_\parallel|$ or the 3D separation
since the fractional uncertainty of the latter two typically reach unity
for sufficiently close pairs.}

The new wide binary system contains the newly discovered white dwarf
\gaia\ DR2 2486388560866377856 (J2015.5 coordinates 02:12:28.34, $-$08:04:17.9), 
which is located at a distance $16.7$\,pc.
The binarity of this system (LP\,649$-$66/67) was already recognised \citep{heintz93-1,luyten97},
however the white dwarf nature of LP\,649$-$66 has only now been confirmed with \gaia\ data.
In SDSS imaging, the white dwarf is only partially resolved from LP\,649$-$67
at a separation of 4\,arcsec.

A noteworthy feature of the binary distribution in Fig.~\ref{fig:Dvdiff} is
the negative correlation between $\Delta v_\perp$ and $D_\perp$.
Because the uncertainties in both dimensions are smaller than the points (Table~\ref{tab:wide_bin}),
we interpret the correlation to result from the relative orbital motion between binary components.
The red dashed line in Fig.~\ref{fig:Dvdiff} indicates the maximum allowable $\Delta v_\perp$
for a pair of 1\,\Msun\ stars separated by $D_\perp$, and on circular orbits,
which is a power-law of $\Delta v_\perp \propto D_\perp^{-1/2}$.
This runs approximately parallel to the top of the binary distribution (in log-log space),
indicating that, for systems in the Solar neighbourhood, \gaia\ astrometry
is indeed sensitive to wide binary orbital motions.

A clear outlier within Fig.~\ref{fig:Dvdiff}
is WD\,0743$-$033 and its established companion
171\,Pup\,A, which are separated by $\simeq 13\,400$\,au and with
$\Delta v_\perp \simeq 4.7$\,\kms. At face-value, this velocity difference and projected separation
implies the two objects are gravitationally unbound.
However, 171\,Pup\,A suffers astrometric excess noise of 1.7\,mas, which we attribute
to its unresolved companion \citep{tokovininetal12-1}.
We checked the PPMXL proper motion \citep{roeseretal10-1} for 171\,Pup\,A, finding  
$(\mu_\mathrm{Ra}, \mu_\mathrm{Dec}) = (-276.8\pm2.3, 1679.3\pm2.4)$\,\masy.
Both PPMXL components are significantly different from the \gaia\ values,
but within $1\,\sigma$ of the \gaia\ proper motion for WD\,0743$-$336 (Table~\ref{tab:wide_bin}).
Given the precision of the PPMXL proper motion, we determined an upper-limit to $\Delta v_\perp$
of 0.45\,\kms\ (95\,percent confidence), shown by the hollow point joined to the \gaia\ value
in Fig.~\ref{fig:Dvdiff}.

For similar reasons, the known companion to WD\,2057$-$493 
\citep[WISEA\,J210104.88$-$490626.5,][]{kirkpatricketal16-1}
failed to pass our initial selection criteria due to astrometric excess noise.
The white dwarf has a \gaia\ parallax of $74.89\pm0.04$\,mas, whereas
$65.65\pm0.49$\,mas is found for the companion,
but with an \textsc{astrometric\_excess\_noise} of 2.3\,mas.
Therefore, the fairly liberal cut given by Equation~\eqref{eq:ddparallel} still causes this
system to be missed.
Unlike for WD\,0743$-$033/171\,Pup\,A, the proper motions for this pair are comparable.
Adopting the parallax of the white dwarf, $D_\perp$ and $\Delta v_\perp$ are respectively
inferred to be $858.54\pm 0.46$\,au and $1.091\pm0.086$\,\kms,
indicated by the hollow point in Fig.~\ref{fig:Dvdiff}.
The astrometric excess noise potentially indicates that WISEA\,J210104.88$-$490626.5
is itself an unresolved binary (similarly to 171\,Pup\,A),
in which case the system would be a hierarchical triple.

Overall, we found our methodology to be successful in identifying wide binary systems
for local \gaia\ stars. With the exceptions of WD\,0743$-$033 and WD\,2057$-$493
described above, all other 20\,pc white dwarfs with known companions that ought to be re-detected,
were easily re-identified in Fig.\,\ref{fig:Dvdiff}.
We note that the known companions to WD\,0642$-$166 (Sirius\,B), WD\,1009$-$184,
and WD\,2341$+$322 are not present in DR2 with five-parameter astrometry, and ostensibly
were not recovered in our search.
Similarly, five known white dwarfs in wide binaries, WD\,0208$-$510, WD\,0413$-$077, WD\,0727+482A/B, and WD\,0736+053 (Table~\ref{tab:notInDR2})
are themselves absent from DR2 (with the exception of Procyon, their companions \emph{are} present),
and so are also not recovered.
We also acknowledge that our methodology is not free of limitations. The described method 
runs in $O\big(N^2\big)$-time for $N$ stars, of which some fraction are white dwarfs.
Thus, in terms of search radius, $r$, the search time scales as $O\big(r^6\big)$.
Furthermore, astrometric excess noise can prohibit the identification of some pairs
if the astrometry for either component are strongly affected.
This is problematic for identifying hierarchical multiples where one
component is an unresolved binary, as was found already for WD\,0743$-$033/171\,Pup.
One final limitation is that our method depends on the high astrometric precision found
predominantly for local white dwarfs. In Table~\ref{tab:wide_bin} we typically measured
$\Delta v_\perp$ to a precision of a few 0.01\,\kms. Beyond 100\,pc
where many white dwarfs have $G\sim 20$, an uncertainty of a several \kms\ can be expected
instead. In such cases, the standard approach of proper motion comparison would prove
more appropriate.

In summary we have identified 23 wide pairs containing white dwarfs within the \gaia\ DR2
data. Of these 23, 21 are wide white dwarf plus main-sequence star (WD+MS) binaries
with two containing newly identified white dwarfs. Two of these WD+MS systems were not directly recovered
by our method because of high astrometric excess noise of the companions,
although are included in Fig.~\ref{fig:Dvdiff} and Table~\ref{tab:wide_bin} for completeness.
For six previously known WD+MS binaries, either the
white dwarf or the main-sequence star is missing from \gaia\ DR2.
Additionally, we recovered two known spatially resolved white dwarf plus white dwarf (WD+WD) 
binaries while one known system is missing from \gaia\ DR2. 
The binary population synthesis from \citet{toonenetal17-1} 
predicts numbers of resolved WD+MS and WD+WD systems that are in the range
23--43 and 16--30, respectively.
We note that these ranges have been updated to account for the higher spatial resolution
of \gaia\ than that of the ground-based observations assumed in \citet{toonenetal17-1}.
These numbers can decrease by 15--30\,percent if we consider
the possible disruption of weakly bound binaries by Galactic interactions.
The lack of wide WD+WD binaries compared to the population model found 
by \citet{toonenetal17-1} based on the ground-based 20\,pc sample is confirmed
with the significantly improved quality of the \gaia\ observations.

\subsection{Unresolved double degenerate candidates}

The \gaia\ sample includes one confirmed close double degenerate, WD\,0135$-$052 \citep{toonenetal17-1} and six white dwarfs, WD\,0121$-$429, WD\,0233$-$242, WD\,0503$-$174, WD\,0839$-$327, WD\,2048+263, and WD\,2248+293, previously described as double degenerate candidates in \citet{giammicheleetal12-1}. \gaia\ does not resolve any of these objects, and in all but two cases our \gaia\ atmospheric parameters (Table~\ref{tab:wd20_gaia}) using a single star model suggest a very low surface gravity inconsistent with single star evolution. The two exceptions are  WD\,0503$-$174 and WD\,0839$-$327 which are in the low surface gravity tail of our distribution in Fig.~\ref{parameters3} but do not stand out as particular outliers. Therefore, \gaia\ data alone is insufficient to confirm the nature of these two objects and spectroscopy is needed for a full diagnostic. From our \gaia\ atmospheric parameters we do not identify any new double degenerate candidates, especially when considering the trend of systematically lower surface gravities at cool \Teff\ in Fig.~\ref{parameters3}. Of the new systems, we found that \gaia\ DR2 2486388560866377856 has a $\log g = 7.67$ from our atmospheric fit. However, the star appears only 4\,arcsec away from its much brighter companion, and thus its photometry may have been contaminated.

The updated models (accounting for the enhanced resolution of \gaia) of \citet{toonenetal17-1}
upredict 0.5--7.0 unresolved double white dwarfs within 20\,pc and therefore the population
synthesis model remains in agreement with the observations.

\subsection{Unresolved main-sequence + white dwarf binaries}

We do not identify any
unresolved white dwarf plus  main-sequence binaries in the 20\,pc \gaia\ sample.
The two previously known systems,
WD\,0419$-$487 and WD\,0454+620 
\citep{toonenetal17-1} are confirmed (Table~\ref{tab:out20}) or suspected (Table~\ref{tab:notInDR2})
to be beyond 20\,pc, respectively. 
Since \citet{toonenetal17-1}, we became aware of two very likely unresolved 
white dwarf plus main-sequence binaries: 

\textbf{G\,203--47ab} was identified by \citet{reid+gizis97-1} as a single-lined
binary with a M3.5 dwarf primary, and \citet{delfosseetal99-1} determined an
orbital period of $14.7136\pm0.005$\,d and a low eccentricity,
$e=0.068\pm0.004$. Adopting a mass of 0.2\,\Msun\ for the M3.5 primary,
\citet{delfosseetal99-1} derived a lower limit on the mass of the companion of
0.5\,\Msun, implying that it has to be a degenerate stellar remnant. Based on
the detected $U-B$ colour excess of G203--47ab, \citet{delfosseetal99-1} argue
that it is most likely a white dwarf. The close proximity of this system,
$D=7.43\pm0.03$\,pc, affects the claimed completeness of the 13\,pc sample. The white
dwarf nature could possibly be confirmed with ultraviolet observations (G203--47ab is
detected in the \textit{GALEX} survey with $FUV=21.85$\,mag and
$NUV=19.58$\,mag). This system is found in DR2 with a source ID 1355264565043431040
and parallax $134.60\pm0.49$\,mas.

\textbf{Wolf\,1130AB} is a single-lined binary with an
orbital period of 0.4967\,d, found by \cite{maceetal18-1} at a distance of $16.7\pm0.2$\,pc.
The primary star is an M-subdwarf with a mass of $\simeq0.3$\,\Msun, which is tidally locked, as
established from the photometric $V$-band modulation. The \textit{Gaia} DR1
parallax and projected rotational velocity were combined to tightly constrain
the inclination, and hence the mass of the unseen companion, Wolf\,1130B, to
$1.24^{+0.19}_{-0.15}$\,\Msun.
The DR2 parallax of this system is $60.391\pm0.034$\,mas ($D=16.550\pm0.009$\,pc) with DR2
source ID 2185716209919937664.

Just as ground-based observations, \gaia\ is unable to detect faint white dwarfs outshone
by bright companions in the optical hence our sample may be incomplete in this respect. 
Using different theoretical assumptions regarding binary evolution and Galactic stellar
formation history, \citet{toonenetal17-1} predict 0.5--1.6 unresolved white dwarf plus
main sequence binaries within 20\,pc at the enhanced resolution of \gaia.
While \gaia\ DR2 can not easily detect new binaries of these types,
the small predicted number suggests it does not significantly impact the estimation of
the white dwarf space-density. Binaries sufficiently wide to produce an
astrometric perturbation, such as suspected for G\,203$-$47ab \citep{delfosseetal99-1}, 
may be identified by future \gaia\ data releases.

\section{Space-density}
\label{sec:spacedensity}

Despite the precision of \gaia\ parallaxes and the new white dwarf identifications,
estimating the white dwarf space-density from our
\gaia\ DR2 sample is not trivial. Of the known degenerate stars,
some of those within (or thought to be within) 20\,pc
do not have \gaia\ 5-parameter astrometric solutions. This
primarily affects the closest systems with large proper motions,
but also a few objects between 10 and 20\,pc.
To accurately calculate the local white dwarf space-density it is thus
important to consider the incompleteness of this sample, and how 
this varies as a function of distance.

We assumed that the \gaia\ 20\,pc
white dwarfs are drawn from a selection function, $S(\vec\theta, D)$, 
representing the probability of detection as a function of distance,\footnotemark\
$D$, and where $\vec\theta$ is the parameter vector defining $S$ (whose shape we
would like to estimate). With a prescription of $S$, we then simply look at how
previously known white dwarfs are re-detected in \gaia\ DR2 as a function of distance.
To constrain $S$, we chose the sample of white dwarfs which were previously estimated to
be within $3\,\sigma$ of $20$\,pc, \emph{and} where a parallax was available before DR2.
We enforced the requirement of prior parallaxes because many of the white dwarfs with
only spectroscopic distances turned out to be predominantly much
farther away than expected upon release of DR2 data (Table~\ref{tab:out20}).
Furthermore, objects where previous parallaxes suggested potential 20\,pc membership,
but are now known to be beyond 20\,pc were still included -- their exclusion would
bias the sample towards non-detections near 20\,pc, and their inclusion
only acts to further constrain $S$ slightly beyond 20\,pc.
For the re-detected objects we used the updated distances from their new \gaia\ parallaxes.
One final point of note, is that we counted both members of wide double degenerates
as a single detection, since their \gaia\ detection/non-detection are not independent.

\footnotetext{While this selection function can be explored in terms of
proper motion, magnitude, or sky-position, etc., for a space-density estimate we
are required to work with distances.}

As re-detection is a binary-categorical variable,
estimation of $\vec\theta$ is naturally suited to logistic regression.
We define $S$ in terms of the logistic function,\footnotemark\
\begin{equation}
	S(\vec\theta, D) = \frac{1}{1+\exp(-f(\vec\theta, D))}~,
    \label{eq:logit}
\end{equation}
where we define $f(\vec\theta, D) = \theta_0 \log(D/\mathrm{pc})+\theta_1$, i.e. a first
order polynomial in the natural logarithm of $D$.

\footnotetext{With this formulation of $S$, $f(\vec\theta, D)$ is equivalent
to the log-odds-ratio, mapping $\mathbb{R}$ to the interval [0, 1].}

Regardless of the detailed formulation of Equation\,\eqref{eq:logit}, the likelihood
of a re-detection for a single white dwarf is a Bernoulli trial with probability $S$.
For a set of detections and non-detections, their combined likelihood is therefore given by
\begin{equation}
	L = P(\vec{y}|\vec\theta, \vec{D}) = \prod_i^N S_i^{y_i} (1-S_i)^{1-y_i}~,
\end{equation}
where $y_i=1$ for a detection, $y_i=0$ for a non-detection, and $S_i = S(\vec\theta, D_i)$.

To estimate the values of $\vec\theta$ and their
uncertainties, we used the MCMC ensemble sampler \textsc{emcee} \citep{foremanmackeyetal13-1}
to sample the log-likelihood,\footnotemark\ which can be simplified to 
\begin{equation}
	\log{L} = \sum_{\substack{i\\y_i=1}} \log(S_i)
            + \sum_{\substack{i\\y_i=0}}\log(1-S_i)~.
\end{equation}
\footnotetext{From a Bayesian perspective, our intention is ultimately
to estimate $P(\vec\theta|D_i, y_i)$, the posterior distribution
on $\vec\theta$. Lacking any informed priors on $\vec\theta$, our sampling of
the log-likelihood, is equivalent to assuming improper uniform priors on both
$\vec\theta$ components.}

\begin{figure}
	\includegraphics[width=\columnwidth]{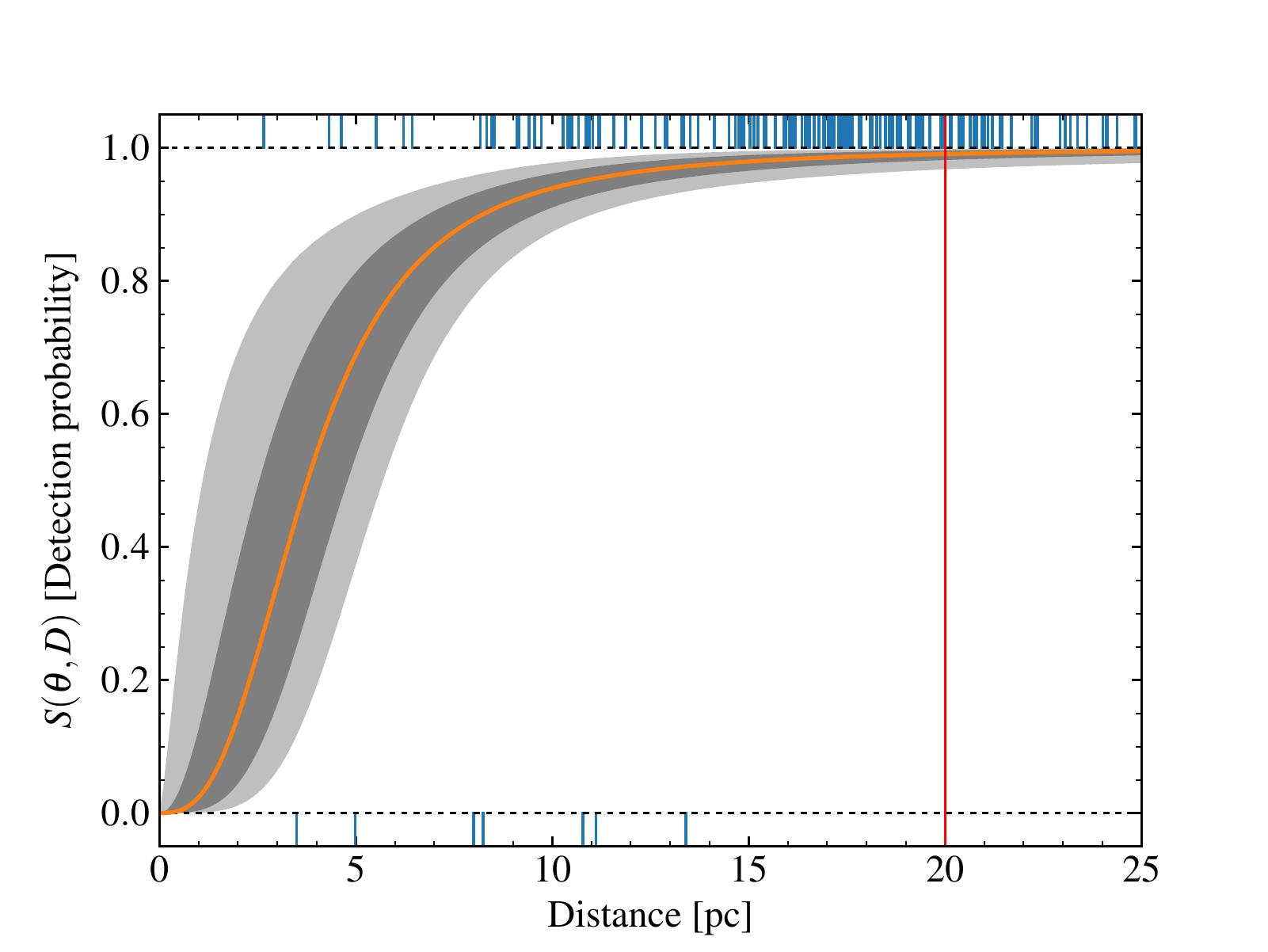}
    \caption{Our fit to the selection function, $S(\vec\theta, D)$.
    The grey regions represent the $1$ and $2\,\sigma$ contours on
    the fit, with the median given in orange.
    The blue lines indicate the locations of detections vs. non-detections for known
    white dwarfs within 20\,pc (to within $3\,\sigma$) prior to DR2.
    }
    \label{fig:scurve}
\end{figure}

From the results of our MCMC, we found $\theta_0$ and $\theta_1$ were
$2.83\pm0.80$ and $-3.7\pm1.8$ respectively,
with a correlation of $-0.969$ between the two.
The corresponding selection function is shown in
Fig.~\ref{fig:scurve} (orange), with its $1\,\sigma$ and $2\,\sigma$
error-contours in grey. This is clearly much better constrained towards higher
distances, due to the increased number of stars per unit distance.
At the distance of Sirius B this corresponds to a detection probability of
$28^{+24}_{-17}$\,percent, but at 20\,pc, this reaches $99.1^{+0.5}_{-1.0}$\,percent.
Integrating over $S$ out to 20\,pc, we determined the effective volume probed by \gaia\,
given by
\begin{equation}
	V_\mathrm{eff} = 4\pi \int_0^{20} S(\vec\theta, D)\, D^2 \,\dd D~.
\end{equation}
Using the MCMC samples for $\vec\theta$ gave
$V_\mathrm{eff} = 3\,2170^{+420}_{-540}$\,\cpc\ implying
a volume-averaged detection-efficiency of \completeness\ out to 20\,pc for \gaia\ white dwarfs.

We calculated the local white dwarf space-density by propagating the uncertainties
in all relevant quantities via a Monte-Carlo approach.
While our DR2 sample contains $139$ sources (Table~\ref{tab:wd20_gaia}),
two systems have parallaxes that could allow them to reside either side of the 20\,pc
boundary. We found probabilities of $\simeq 2/46/52$\,percent for $137/138/139$ systems
within 20\,pc respectively, which we used for our Monte-Carlo draws.
Additionally WD\,0135$-$052 is known to be an unresolved double degenerate system,
therefore we added one to the above quantity to obtain the total white dwarf count, $N_\mathrm{Gaia}$.

\begin{figure}
	\includegraphics[width=\columnwidth]{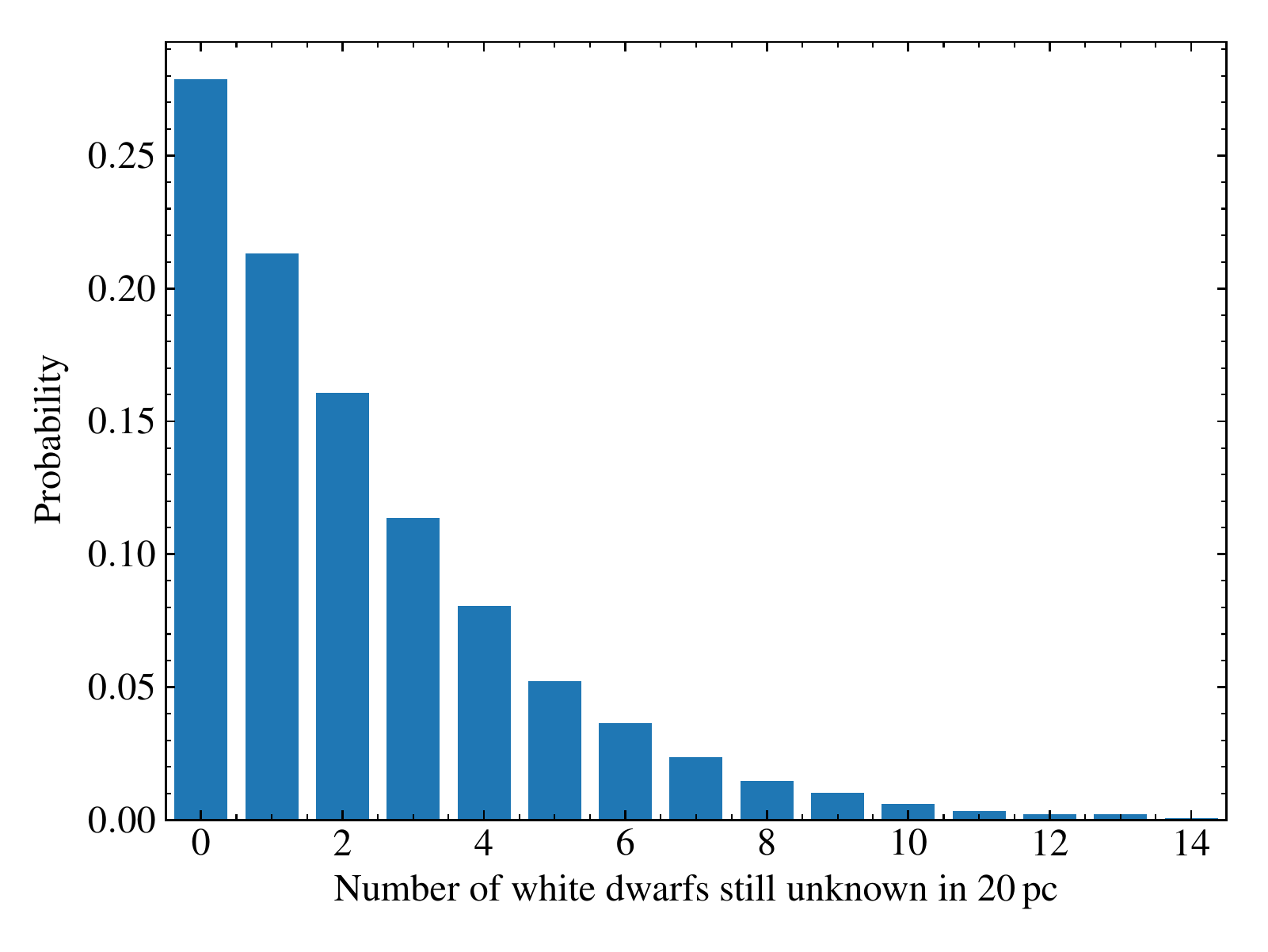}
    \caption{Probability distribution for the number of 20\,pc white dwarfs that
    remain undetected both in \gaia\ and elsewhere.}
    \label{fig:nmissing}
\end{figure}

Accurately ascertaining the space-density required that we also constrain the number of missing
white dwarfs ($N_\mathrm{missing}$) within 20\,pc.
Using our Monte-Carlo samples for both $N_\mathrm{Gaia}$ and our
estimate of the completeness (as calculated above) we determined $N_\mathrm{missing}$
using the negative-binomial distribution, which we found had a mode of five.
However, eight white dwarfs that are confirmed 20\,pc members are missing from our 20\,pc sample
(Table~\ref{tab:notInDR2}), providing a prior on the number of missing systems.
Thus, we discarded all samples with values less than eight, with the remaining samples
used for our estimate of $N_\mathrm{missing}$. 
Our calculation for the total number of 20\,pc white dwarfs is therefore given by
$N_\mathrm{tot} = N_\mathrm{Gaia} + N_\mathrm{missing}$.
As a brief aside, $N_\mathrm{missing}-8$ estimates the remaining number of 20\,pc white dwarfs
that are yet to be discovered (Fig.~\ref{fig:nmissing}), which has a mode of zero
at 28\,percent probability, a median of two, and a 95\,percent upper limit of seven.
Correspondingly the completeness for \emph{all} white dwarfs within 20\,pc (not just those in
\gaia) has a median value of
98.7\,percent, with 95.5\,percent as a lower-limit (95\,percent confidence).
Therefore, there is a reasonable chance that all 20\,pc white dwarfs have now
been identified, though the prospect of several more members
remains for future \gaia\ data releases.

\begin{figure}
	\includegraphics[width=\columnwidth]{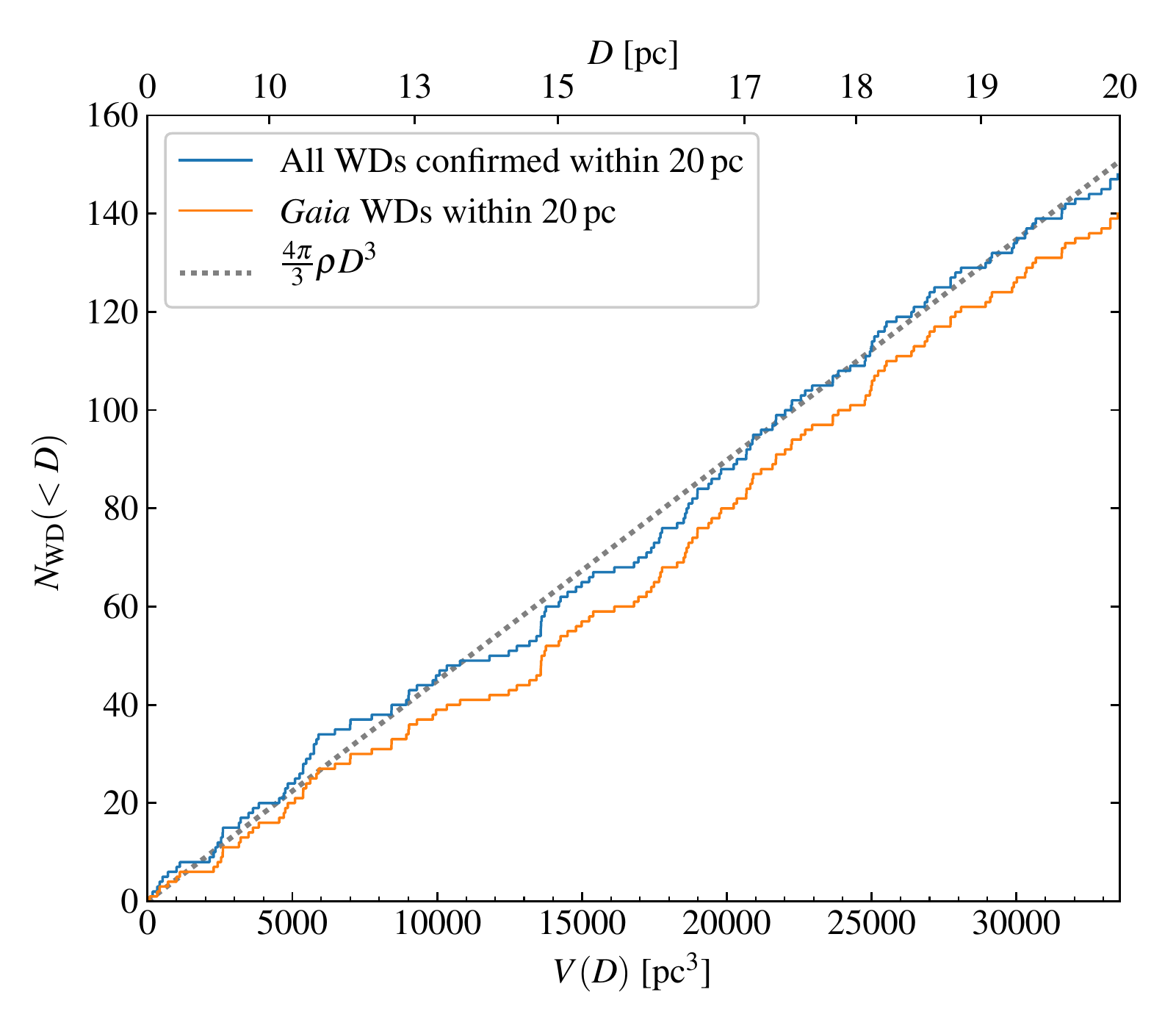}
    \caption{Cumulative number of all-known and \gaia\ white dwarfs
    as a function of enclosed volume (blue and orange respectively).
    The grey dotted line shows the expectation value at each distance for the adopted
    space-density. 
    Note that the slope of this line depends only on the total number
    of objects enclosed within 20\,pc, and does not represent a fit
    to the cumulative distribution.
    Fitting directly to the cumulative distribution
    is erroneous as the steps are not statistically independent,
    and leads to an underestimated uncertainty in the space-density.}
    \label{fig:n3}
\end{figure}

We considered $N_\mathrm{tot}$ as drawn from a Poisson-process with mean $\bar{N}$.
To correctly account for the Poisson-uncertainty associated with $N_\mathrm{tot}$,
we drew samples from a Gamma-distribution (the conjugate prior of the Poisson distribution)
with a Jeffreys prior\footnotemark\ to constrain $\bar{N}$, finding $\bar{N} = 150.3\pm12.6$.
Finally, by dividing by the volume of the entire
20\,pc sphere, we arrived at our adopted space-density, $\rho = \WDdensity$.
In Fig.~\ref{fig:n3} we show the cumulative distribution for our \gaia\ sample compared
with the expectation value from our space-density calculation.

While slightly lower than some other recent values, our adopted space-density is within $1\,\sigma$ of
these other estimates calculated from the local sample -- $(4.8\pm0.5)\times10^{-3}$\,\pcpc\,was
reported by \citet{holbergetal16-1}, $4.39\times10^{-3}$\,\pcpc\,by \citet{giammicheleetal12-1},
and $(4.9\pm0.5)\times10^{-3}$\,\pcpc\,by \citet{sionetal09-1}.
Such close agreement should not come as a surprise as these samples necessarily contain some
overlap, and are thus not statistically independent measurements.
On the other hand, \citet{munnetal17-1} made an independent measurement of
the space-density using $\simeq 3000$ SDSS white dwarfs finding
$\rho=5.5\pm0.1$\,\pcpc. While this value is in disagreement with our estimate
at the $2.6\,\sigma$ level, we note that the uncertainty in the 
density adopted by \citet{munnetal17-1} appears to be entirely statistical and does not, for instance,
include possible uncertainty in the Galactic model used in their calculation.
A fractional systematic uncertainty of 10\,percent would be sufficient to resolve
this discrepancy.

\footnotetext{The Jeffreys prior for the rate parameter, $\lambda$, of the Poisson distribution is
$P(\lambda) \propto \lambda^{-1/2} \equiv \mathrm{Gamma}(1/2, 0)$.}

Further improvement to the precision of the local space-density will necessitate
a vast increase in the number of objects, and thus an extension of this work to a larger volume.
We do note however that our estimate for $\rho$ does not consider white dwarfs hidden
as close companions to bright main-sequence stars.

\section{Summary}
\label{sec:summary}

We have revisited the 20\,pc white dwarf local sample making use of the recently released
\gaia\ DR2 data.
The updated sample is now potentially complete for resolved sources, even for the most
intrinsically faint white dwarfs. Within 20\,pc, we identified 130 known white dwarfs, while having confirmed
57 candidate 20\,pc members (but known white dwarfs) to be located at farther distances.
From our Hertzsprung-Russell diagram, we identified 15 new 20\,pc white dwarf candidates.
Fitting not only the \gaia\ data, but also, where available, Pan-STARRS and 2MASS photometry,
we established 9 of these to be newly identified white dwarfs (with the closest system
found at $13.05$\,pc), with the remainder as
likely main-sequence objects with adversely affected \gaia\ photometry/astrometry.
Despite the high quality of the \gaia\ observations of very nearby stars, we therefore stress
significant care has to be taken in the sample selection and data analysis to avoid misinterpreting
the relatively large fraction of spurious sources.
Spectroscopic follow-up observations of the newly identified white dwarfs 
and white dwarf candidates are encouraged to determine their photospheric properties.
The rich diversity of spectral types and binary evolution paths observed in the currently characterised
local sample \citep{giammicheleetal12-1, toonenetal17-1} means that it is a
benchmark to understand stellar and binary evolution as well as the spectral evolution
of the atmosphere of white dwarfs owing to carbon dredge-up, convective mixing,
and the accretion of planetesimals \citep{dufouretal05-1,tremblay+bergeron08-1,gentileetal17-1,hollandsetal18-1}.

Making full use of the exquisite \gaia\ astrometry, we searched for wide companions to
the white dwarfs in the 20\,pc sample. For each white dwarf in the sample, we considered all
stars within 1\,pc, and compared tangential velocity differences with the projected separation
to distinguish companions. Of the known wide binaries containing white dwarfs where both
components were found in DR2, all but two were easily re-identified (17 WD+MS and 2 WD+WD binaries).
The companions to the other two white dwarfs either had an inaccurate proper motion or parallax
owing to their high astrometric excess noise. Of the nine new white dwarfs, we found one
of these has a wide M dwarf companion. We did not identify any companions to known
white dwarfs that were hitherto undiscovered. Since no new resolved WD+WD systems were identified,
the total number of wide double degenerate systems within 20\,pc of the Sun remains at three,
including one pair absent from the \gaia\ astrometry. This observational finding remains in contrast
with the most recent binary population synthesis models which instead predict 16--30 such systems.
In terms of unresolved binaries, we did not identify any new systems. Candidate systems 
identified elsewhere remain as such, with none being partially resolved into two components.
High-resolution, time-resolved spectroscopic follow-up will be required to confirm their
binarity one way or the other.
In any case, the number of confirmed unresolved binaries (both WD+WD and WD+MS systems) remain
consistent with the expected values from binary population models.

With this up-to-date sample in hand we calculated a new estimate of the local white dwarf space density.
To be as accurate as possible in this calculation, we placed statistical constraints on the number
of white dwarfs remaining undetected within 20\,pc. For this purpose, we first assessed the re-detection
rate of known white dwarfs as a function of distance, and found that while \gaia\ does not
reliably detect white dwarfs at the very closest distances, the detection
probability is close to 100\,percent at 20\,pc.
This implied that white dwarfs are found by \gaia\ with an overall
detection efficiency of \completeness\ within the 20\,pc volume.
Considering 8 degenerate stars that did not have \gaia\ DR2 and that have been firmly established elsewhere
as 20\,pc members, we found an upper limit of 7 for the number of 20\,pc white dwarfs remaining to
be found. Folding this uncertainty into our calculations, we determined
the local white dwarf space density to be \WDdensity, which we found to be consistent with,
but more precise than previous values determined from local white dwarfs.
Because we find the 20\,pc sample is now close to complete,
more precise estimates of the local white dwarf space-density will necessitate extending
this work to larger volume-complete samples.

While we have successfully assessed many properties of the local \gaia\
white dwarf population, this represents only a subset of white dwarf
related science that can be probed with \gaia.
For example, the statistics of local white dwarfs with magnetic fields or metal-pollution 
were considered beyond the scope of this work,
but represent potential avenues of future investigation using \gaia\ data.
More generally, magnitude-limited \gaia\ white dwarf samples will
naturally contain several orders of magnitude more stars than any local sample,
but at the expense of increased selection bias.
We therefore expect this work to be useful as a comparison sample in these cases.
Future \gaia\ data releases will allow the extension of the volume-complete
local sample to greater distances as an increased number of faint objects acquire full
5-parameter astrometry.

\section*{Acknowledgements}
The research leading to these results has received funding from the European
Research Council under the European Union's Seventh Framework Programme
(FP/2007- 2013) / ERC Grant Agreement n. 320964 (WDTracer) and under the
European Union's Horizon 2020 research and innovation programme n. 677706
(WD3D).




\bibliographystyle{mnras}
\bibliography{aamnem99,aabib} 




\appendix
\onecolumn
\newpage

\setlength{\tabcolsep}{0.5ex}

\scriptsize
\begin{landscape}
\begin{longtable}{llcccccccccccc}
	\caption{\label{tab:wd20_gaia} Astrometric and photometric data for our
    \gaia\ 20\,pc white dwarf sample, as well as the results
    from our model atmosphere fits from Section~\ref{sec:atmfit}. The column comp.
    refers to the adopted atmospheric composition used for our fits to the
    \gaia\ photometry and parallaxes. Note that the coordinates are given
    in the \gaia\ DR2 J2015.5 epoch, and so must be corrected by the
    proper motions for direct comparison with the J2000 epoch. The parallaxes
    are given without the 0.03\,mas offset suggested by \citet{gaiaDR2-collab-2}.
    } \\
    \hline
    J2015.5 & WD & \gaia\ DR2 ID & $\varpi$ [mas] & $\mu_\mathrm{Ra}$ [\masy] &
    $\mu_\mathrm{Dec}$ [\masy] & $G$ [mag] & \bp\ [mag] & \rp\ [mag] & spType & comp. &
    \Teff [K] & $\log g$ [dex] & Note \\
    \hline
    \endfirsthead
	\caption{continued.} \\
    \hline
    J2015.5 & WD & \gaia\ DR2 ID & $\varpi$ [mas] & $\mu_\mathrm{Ra}$ [\masy] &
    $\mu_\mathrm{Dec}$ [\masy] & $G$ [mag] & \bp\ [mag] & \rp\ [mag] & spType & comp. &
    \Teff [K] & $\log g$ [dex] & Note \\
    \hline
    \endhead
    \hline
    \endfoot
    \hline \\[-4mm]
    \caption*{Notes:
    (1)~For our model atmosphere fits, $\log[\mathrm{H/He}] = -1$ and $-3$ were assumed
    for WD\,0038$-$226 and WD\,2008$-$600 respectively.
    (2)~Newly identified white dwarf.
    (3)~The \textsc{phot\_bp\_rp\_excess\_factor} value of 1.661 indicates that the
        modelling of the background flux is not good enough to fit the \gaia\ fluxes.
    (4)~Known white dwarf, but not previously considered a potential 20\,pc member.
    (5)~White dwarf candidate recently identified by \citet{smithetal18-1} via its
        proper-motion.}
    \endlastfoot
    000211.59$-$431006.0 & 2359$-$434   & 4994877094997259264 & $119.98\pm0.03$ & $ +613.64\pm0.04$ & $ -686.73\pm0.04$ & $12.9137\pm0.0004$ & $12.9943\pm0.0034$ & $12.7015\pm0.0009$ & DAP & H & $ 8464\pm 21$ & $8.36\pm0.01$ & \\
000240.29$-$341351.4 & 0000$-$345   & 2312821266302222720 & $ 67.54\pm0.04$ & $ +152.20\pm0.06$ & $ -762.18\pm0.06$ & $14.8943\pm0.0005$ & $15.1187\pm0.0040$ & $14.4948\pm0.0020$ & DC & He & $ 6327\pm 29$ & $8.14\pm0.01$ & \\
000720.92$+$123017.6 & 0004$+$122   & 2766234439302571904 & $ 57.31\pm0.11$ & $ +354.94\pm0.18$ & $ -232.26\pm0.11$ & $16.2650\pm0.0008$ & $16.6828\pm0.0062$ & $15.6775\pm0.0032$ & DC & He & $ 5116\pm 24$ & $8.21\pm0.02$ & \\
001214.00$+$502512.2 & 0009$+$501   & \phantom{0}395234439752169344 & $ 91.98\pm0.03$ & $ -461.37\pm0.04$ & $ -551.63\pm0.02$ & $14.2474\pm0.0002$ & $14.4704\pm0.0020$ & $13.8494\pm0.0009$ & DAH & H & $ 6476\pm 11$ & $8.24\pm0.01$ & \\
001350.66$-$714958.1 & 0011$-$721   & 4689789625044431616 & $ 53.23\pm0.03$ & $ +233.03\pm0.04$ & $ -247.70\pm0.04$ & $15.0466\pm0.0005$ & $15.2958\pm0.0025$ & $14.6254\pm0.0017$ & DA & H & $ 6271\pm 18$ & $7.87\pm0.01$ & \\
001412.21$-$131112.2 & 0011$-$134   & 2418116963320446720 & $ 53.81\pm0.07$ & $ -546.70\pm0.10$ & $ -714.86\pm0.08$ & $15.7657\pm0.0006$ & $16.0738\pm0.0035$ & $15.2799\pm0.0027$ & DAH & H & $ 5858\pm 22$ & $8.20\pm0.01$ & \\
004125.51$-$222108.2 & 0038$-$226   & 2349916559152267008 & $109.88\pm0.03$ & $ -463.23\pm0.04$ & $ -381.77\pm0.03$ & $14.3074\pm0.0002$ & $14.6778\pm0.0013$ & $13.7647\pm0.0010$ & DQpec & H+He & $ 5162\pm 12$ & $7.86\pm0.01$ & 1\\
004911.18$+$052237.0 & 0046$+$051   & 2552928187080872832 & $231.74\pm0.04$ & $+1231.33\pm0.08$ & $-2711.83\pm0.04$ & $12.3122\pm0.0001$ & $12.5531\pm0.0006$ & $11.9427\pm0.0005$ & DZ & He & $ 6424\pm  5$ & $8.25\pm0.01$ & \\
\end{longtable}
\end{landscape}
\normalsize

\newpage
\small
\begin{longtable}{lccccccc}
	\caption{\label{tab:out20} White dwarfs previously thought to reside within
    20\,pc that are now confirmed to be outside the 20\,pc sample. The reference
    column indicates the source of the previous parallax/distance estimate. The
    0.03\,mas offset \citep{gaiaDR2-collab-2} is not included in $\varpi_\mathrm{DR2}$,
    but \emph{is} folded into the calculation of $D_\mathrm{DR2}$.
    } \\
    \hline
    J2015.5 & WD & \gaia\ DR2 ID & $\varpi_\mathrm{DR2}$ [mas] & $D_\mathrm{DR2}$ [pc] &
    $\varpi_\mathrm{old}$ [mas] & $D_\mathrm{old}$ [pc] & Ref. \\
    \hline
    \endfirsthead
	\caption{continued.} \\
    \hline
    J2015.5 & WD & \gaia\ DR2 ID & $\varpi_\mathrm{DR2}$ [mas] & $D_\mathrm{DR2}$ [pc] &
    $\varpi_\mathrm{old}$ [mas] & $D_\mathrm{old}$ [pc] & Ref. \\
    \hline
    \endhead
    \hline
    \endfoot
    \hline \\[-4mm]
    \caption*{References:
    (1)~\citet{limogesetal15-1},
    (2)~\citet{gliese+jahreiss91-1},
    (3)~\citet{gianninasetal11-1},
    (4)~\citet{vanaltenaetal95-1},
    (5)~\citet{subasavageetal17-1},
    (6)~\citet{finchetal18-1},
    (7)~\citet{holbergetal16-1}
    (8)~\citet{subasavageetal09-1},
    (9)~\citet{subasavageetal08-1},
    (10)~\citet{vanleeuwen07-1},
    (11)~\citet{lepineetal09-1}.
    }
    \endlastfoot
    000754.60$+$394731.0 & 0005$+$395 & \phantom{0}383108338321272448 & $29.038\pm0.064$ & $34.402\pm0.075$ &  & $20.2\pm4.2$ & 1 \\
002214.93$+$423638.0 & 0019$+$423 & \phantom{0}385105360675267840 & $29.318\pm0.060$ & $34.074\pm0.070$ & $42\pm4$ & $23.8\pm2.3$ & 2 \\
002640.19$-$552451.1 & 0024$-$556 & 4920057871348614272 & $41.553\pm0.032$ & $24.049\pm0.019$ &  & $19$ & 3 \\
010350.26$+$050434.2 & 0101$+$048 & 2552121179905893888 & $44.864\pm0.119$ & $22.275\pm0.059$ & $46.9\pm3.8$ & $21.3\pm1.7$ & 4 \\
011044.51$+$275813.3 & 0108$+$277 & \phantom{0}307323228064848512 & $26.347\pm0.112$ & $37.912\pm0.161$ &  & $26.6\pm4.4$ & 1 \\
021616.14$+$395123.9 & 0213$+$396 & \phantom{0}332820971434386432 & $49.077\pm0.041$ & $20.364\pm0.017$ & $50.75\pm0.82$ & $19.70\pm0.32$ & 5 \\
021659.23$+$425756.2 & 0213$+$427 & \phantom{0}351429930856438656 & $48.889\pm0.094$ & $20.442\pm0.039$ & $50.2\pm4.1$ & $19.9\pm1.7$ & 4 \\
023407.98$-$051137.9 & 0231$-$054 & 2488960249844340352 & $41.628\pm0.055$ & $24.005\pm0.032$ & $55.0\pm10.4$ & $18.2\pm3.6$ & 6 \\
\end{longtable}
\normalsize

\newpage
\normalsize
\begin{longtable}{lccccccc}
	\caption{\label{tab:rejected} Rejected white dwarf candidates from our selection
    of 20\,pc sources. The parallaxes and proper motions are the \gaia\ measurements
    irrespective of whether they are considered to be correct. AEM is the astrometric excess noise.
    The \Teff\ and $\log g$ are from our fits to the absolute photometry of \gaia\, and where
    available Pan-STARRS and 2MASS, \emph{assuming} a white dwarf with a hydrogen dominated
    atmosphere.} \\
    \hline
    \gaia\ DR2 ID & $\varpi$ [mas] & $\mu_\mathrm{Ra}$ [\masy] & $\mu_\mathrm{Dec}$ [\masy] &
    AEM [mas] & \Teff\ [K] & $\log g$ & Note \\
    \hline
    \endfirsthead
	\caption{continued.} \\
    \hline
    \gaia\ DR2 ID & $\varpi$ [mas] & $\mu_\mathrm{Ra}$ [\masy] & $\mu_\mathrm{Dec}$ [\masy] &
    AEM [mas] & \Teff\ [K] & $\log g$ & Note \\
    \hline
    \endhead
    \hline
    \endfoot
    \hline \\[-4mm]
    \caption*{Notes:
    (1) Atypically large astrometric noise, low proper motion, and infrared photometry
        that is inconsistent with the optical fit.
    (2) Extremely small proper motion and moderate astrometric noise.
    (3) Infrared 2MASS photometry suggests a cool main-sequence object. Furthermore, the low
    PPMXL proper motion value of $(\mu_\mathrm{Ra}, \mu_\mathrm{Dec}) = (-5.2, +3.9)\pm3.8$ \masy\ is inconsistent with \gaia\ DR2.
    (4) Low proper motion, moderate astrometric noise and low measured $\log g$ assuming
    	a white dwarf. The even lower PPMXL proper motion of $(+3.4, -10.3) \pm 4.0$ \masy\ is discrepant with \gaia\ DR2.
    (5) The proper-motion is inconsistent
    	with data from elsewhere calling into question the accuracy of the parallax (see Section~\ref{sec:newWDs}).
        Furthermore this star appears very close to a much brighter B-type star.
    (6) Low proper motion and low $\log g$ required to fit the photometry and parallax. The very low PPMXL proper-motion of $(+4.9, -1.6) \pm 7.3$ \masy\ is inconsistent with {\it Gaia} DR2.
    }
    \endlastfoot
    1474900771698364672 & $ 64.99\pm0.48$ & $  -26.26\pm0.37$ & $  -22.85\pm0.24$ & 0.80 & $ 5357\pm 13$ & $8.14\pm0.01$ & 1\\
3456953931703563264 & $ 68.57\pm0.76$ & $   +1.33\pm0.41$ & $   +0.09\pm0.31$ & 0.48 & $ 4532\pm 18$ & $7.42\pm0.01$ & 2\\
4519850409232889856 & $ 53.54\pm0.28$ & $  -19.61\pm0.21$ & $ -116.96\pm0.48$ & 0.77 & $ 3248\pm285$ & $6.45\pm0.01$ & 3\\
6031231499269624832 & $ 55.72\pm0.23$ & $  +37.66\pm0.53$ & $  +29.80\pm0.21$ & 0.35 & $ 4375\pm  8$ & $6.62\pm0.01$ & 4\\
6206558253342895488 & $ 60.44\pm0.60$ & $  +71.09\pm0.82$ & $  -54.16\pm0.56$ & 0.62 & $ 4852\pm 57$ & $8.59\pm0.01$ & 5\\
6432981674985955584 & $ 59.41\pm0.20$ & $  -39.70\pm0.21$ & $   -2.44\pm0.34$ & 0.36 & $ 5988\pm  5$ & $6.66\pm0.01$ & 6\\
\end{longtable}
\normalsize
\begin{longtable}{llccccc}
	\caption{\label{tab:notInDR2} Known white dwarfs within 20\,pc and
    potential 20\,pc members without reliable five-parameter astrometric
    solutions. This includes systems in binaries where the companion is detected,
    as well as sources with incorrect five-parameter solutions that are omitted
    from the \gaia\ DR2 catalogue.
    Most systems have at least 2-parameter \gaia\ detections (with \gaia\ DR2
    source IDs provided), but are noted where not detected at all.
    The reference column indicates the source of the quoted distance or parallax.} \\
    \hline
    WD & Alt. name & \gaia\ DR2 ID & $\varpi$ [mas] & $D$ [pc] & Ref. & Note \\
    \hline
    \endfirsthead
	\caption{continued.} \\
    \hline
    WD & Alt. name & \gaia\ DR2 ID & $\varpi$ [mas] & $D$ [pc] & Ref. & Note \\
    \hline
    \endhead
    \hline
    \endfoot
    \hline \\[-4mm]
    \caption*{References:
    (1)~Gaia DR2 (companion),
    (2)~\citet{limogesetal15-1},
    (3)~\citet{vanaltenaetal95-1},
    (4)~\citet{vanleeuwen07-1}.
    \smallskip \\
    Notes:
    (a)~White dwarf does not have DR2 five-parameter astrometry, but known companion does.
    (b)~A full \gaia\ astrometric solution is listed, but with a parallax of $11.36\pm0.23$\,mas
        and $\textsc{astrometric\_excess\_noise} > 1$\,mas. We do not rule out that it corresponds
        to this DA + dM system.
    (c)~Three sources of similar magnitude and no \gaia\ astrometry at this position.
    (d)~No \gaia\ detection due to saturation of Procyon\,A.
    (e)~No \gaia\ detection, noting that the white dwarf has a
        $\simeq 4$\,arcsec\,yr$^{-1}$ proper motion.
    (f)~Detected with the established proper motion, but with a parallax of $1.44\pm0.55$\,mas.
    }
    \endlastfoot
    0208$-$510 & GJ\,86\,B & 4937000898854436608 & $92.704\pm0.045$ & $10.787\pm0.005$ & 1 & a\\
0413$-$077 & 40\,Eri\,B & 3195919254110817408 & $199.46\pm0.32$ & $5.014\pm0.008$ & 1 & a\\
0454$+$620 & PM\,J04586$+$6209 & \phantom{0}477633967143589760 &  & $21.6\pm1.2$ & 2 & b\\
0553$+$053 & G\,99$-$47 & 3320184202856027776 & $125.0\pm3.6$ & $8.0\pm0.2$ & 3 & \\
0727$+$482A & G\,107$-$70A & \phantom{0}975968340912004352 & $88.543\pm0.066$ & $11.294\pm0.008$ & 1 & a,\,c\\
0727$+$482B & G\,107$-$70B & \phantom{0}975968340910692736 & $88.543\pm0.066$ & $11.294\pm0.008$ & 1 & a,\,c\\
0736$+$053 & Procyon\,B & \phantom{0000000000000000000} & $284.56\pm1.26$ & $3.514\pm0.016$ & 4 & d\\
1121$+$216 & Ross\,627 & 3978879594463069312 & $74.4\pm2.8$ & $13.4\pm0.5$ & 3 & \\
\end{longtable}
\normalsize

\newpage
\begin{landscape}
\begin{longtable}{lccccccccc}
	\caption{\label{tab:wide_bin} Wide binaries identified in our 20\,pc white dwarf sample.} \\
    \hline
    J2015.5 & \gaia\ DR2 ID & $\varpi$ [mas] & $\mu_\mathrm{Ra}$ [\masy] & $\mu_\mathrm{Dec}$ [\masy] &
    \Gabs\ [mag] & $\bp-\rp$ & $D_\perp$ [au] & $\Delta v_\perp$ [\kms] & Note \\
    \hline
    \endfirsthead
	\caption{continued.} \\
    \hline
    J2015.5 & \gaia\ DR2 ID & $\varpi$ [mas] & $\mu_\mathrm{Ra}$ [\masy] & $\mu_\mathrm{Dec}$ [\masy] &
    \Gabs\ [mag] & $\bp-\rp$ & $D_\perp$ [au] & $\Delta v_\perp$ [\kms] & Note \\
    \hline
    \endhead
    \hline
    \endfoot
    \hline \\[-4mm]
    \caption*{Notes: (1)~Both the white dwarf and companionship are newly identified in this work.
    (2)~Companion astrometry adversely affected by astrometric noise.
    (3)~Wide double-degenerate system.}
    \endlastfoot
    015151.68$+$642549.4 & \phantom{0}518201792978858880 & $ 57.77\pm0.02$ & $ +224.82\pm0.03$ & $ -201.52\pm0.03$ & $ 12.7735\pm 0.0009$ & $ +0.256\pm 0.003$ & \multirow{2}{*}{$ 233.67\pm 0.09$} & \multirow{2}{*}{$1.244\pm0.003$} &  \\
015151.70$+$642602.9 & \phantom{0}518201792978856960 & $ 57.76\pm0.03$ & $ +237.74\pm0.04$ & $ -193.60\pm0.04$ & $  9.2094\pm 0.0012$ & $ +2.345\pm 0.001$ &  &  & \smallskip \\
021228.34$-$080417.9 & 2486388560866377856 & $ 59.88\pm0.05$ & $ -609.29\pm0.10$ & $ -448.00\pm0.08$ & $ 12.6241\pm 0.0019$ & $ +0.349\pm 0.009$ & \multirow{2}{*}{$  60.31\pm 0.05$} & \multirow{2}{*}{$3.242\pm0.012$} & \multirow{2}{*}{1} \\
021228.58$-$080417.4 & 2486388560866377728 & $ 59.83\pm0.08$ & $ -604.92\pm0.17$ & $ -488.69\pm0.14$ & $ 10.0750\pm 0.0031$ & $ +2.367\pm 0.004$ &  &  & \smallskip \\
041629.95$-$591800.0 & 4678664766393827328 & $ 54.48\pm0.05$ & $  -45.10\pm0.08$ & $ -175.25\pm0.09$ & $ 11.1718\pm 0.0021$ & $ -0.139\pm 0.005$ & \multirow{2}{*}{$ 239.34\pm 0.22$} & \multirow{2}{*}{$0.705\pm0.028$} &  \\
041628.93$-$591810.4 & 4678664766394767616 & $ 54.71\pm0.16$ & $  -48.30\pm0.28$ & $ -167.80\pm0.30$ & $  2.7540\pm 0.0072$ & $ +1.214\pm 0.004$ &  &  & \smallskip \\
043115.25$+$585811.1 & \phantom{0}470826482635701376 & $181.28\pm0.05$ & $+1335.04\pm0.08$ & $-1947.63\pm0.09$ & $ 13.6445\pm 0.0007$ & $ +0.434\pm 0.005$ & \multirow{2}{*}{$  56.39\pm 0.01$} & \multirow{2}{*}{$2.656\pm0.043$} &  \\
043114.12$+$585805.8 & \phantom{0}470826482635704064 & $180.42\pm0.59$ & $+1303.27\pm2.05$ & $-2043.85\pm1.23$ & $ 10.9795\pm 0.0074$ & $ +2.913\pm 0.005$ &  &  & \smallskip \\
\end{longtable}
\end{landscape}
\normalsize

\appendix
\section{}
\label{app_A}

\begin{figure*}
\centering
\includegraphics[width=0.23\columnwidth,angle=270,bb = 10 145 260 490]{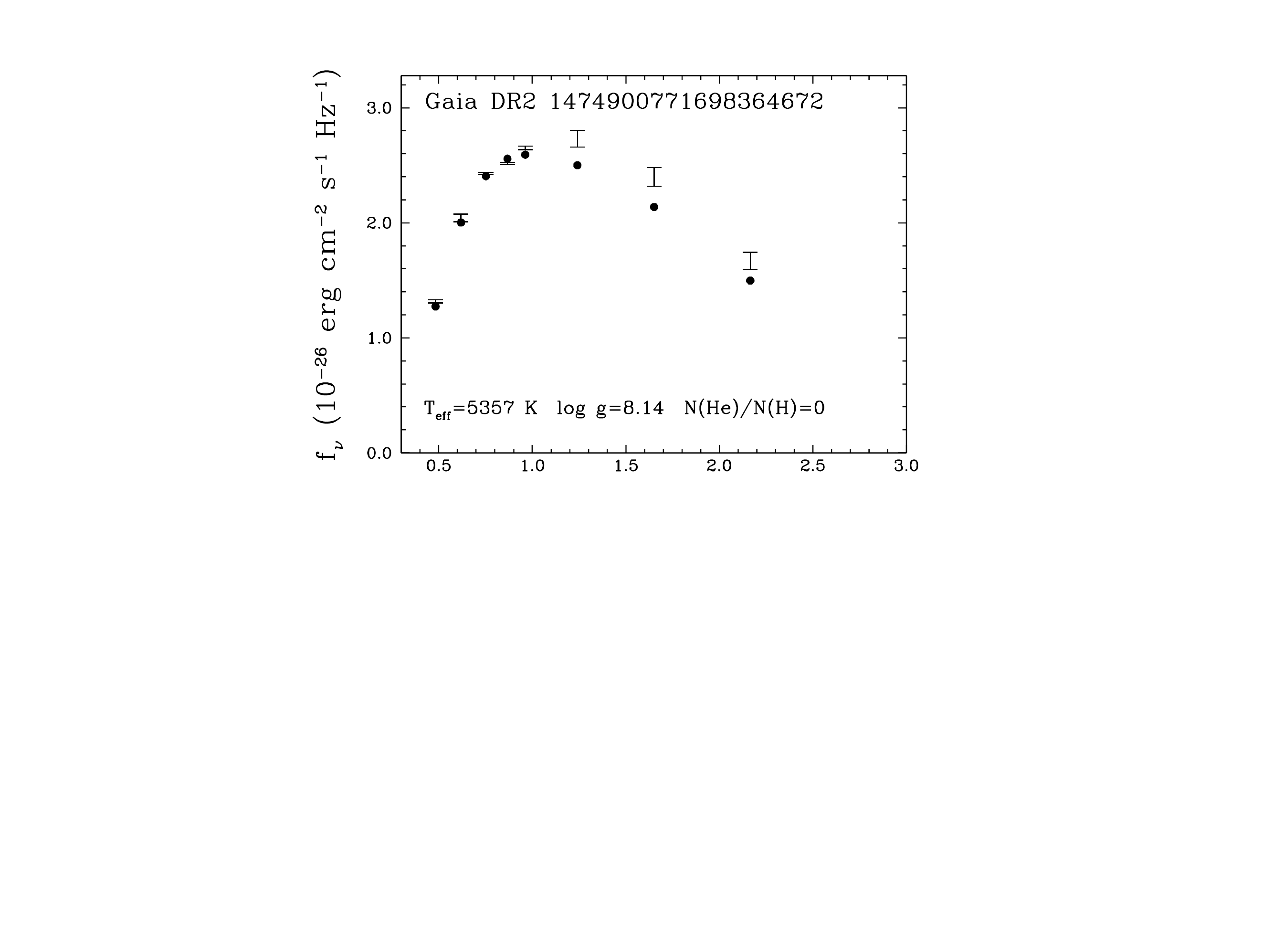}
\includegraphics[width=0.23\columnwidth,angle=270,bb = 10 145 260 490]{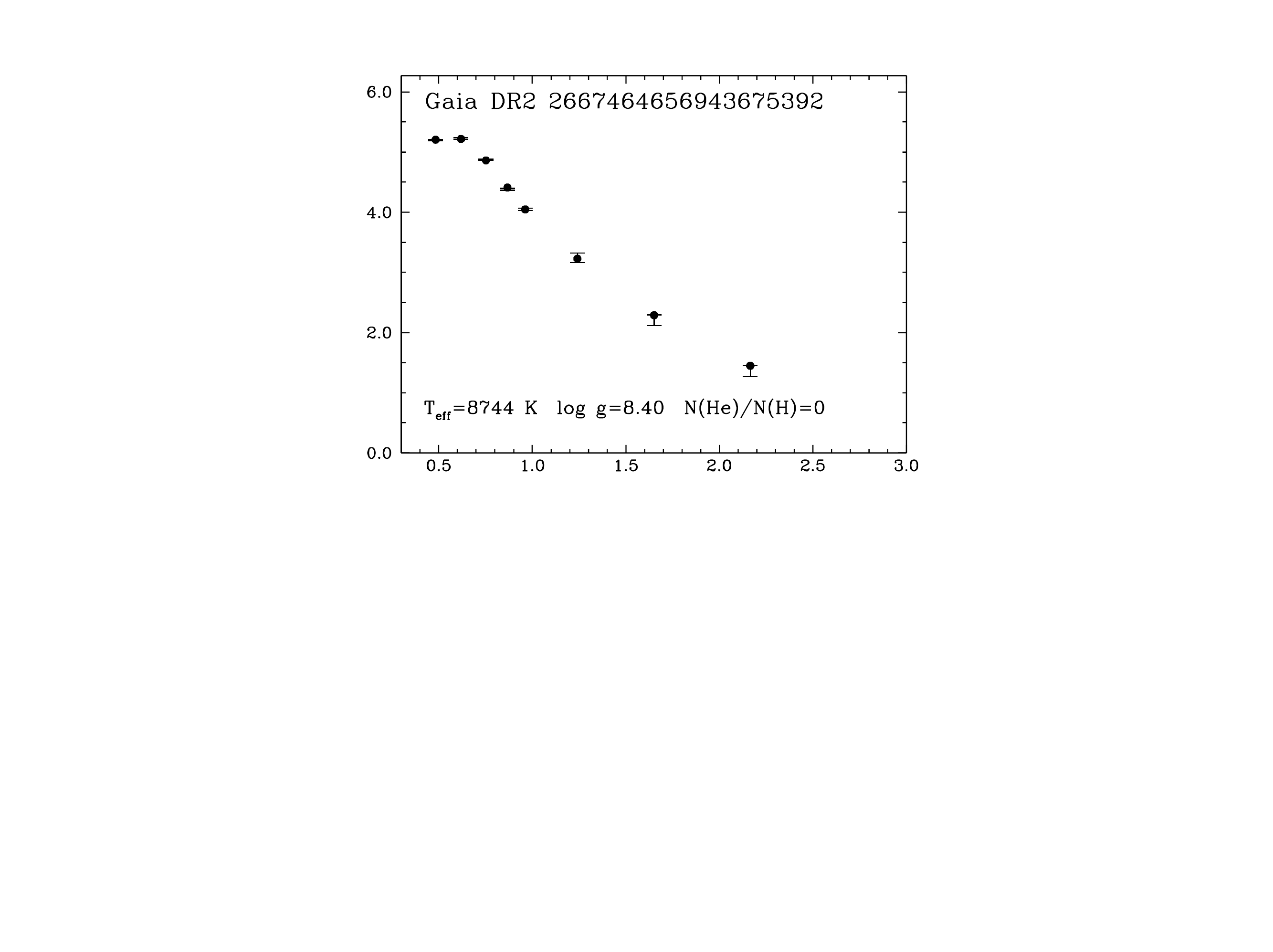}
\newline
\includegraphics[width=0.23\columnwidth,angle=270,bb = 10 145 260 490]{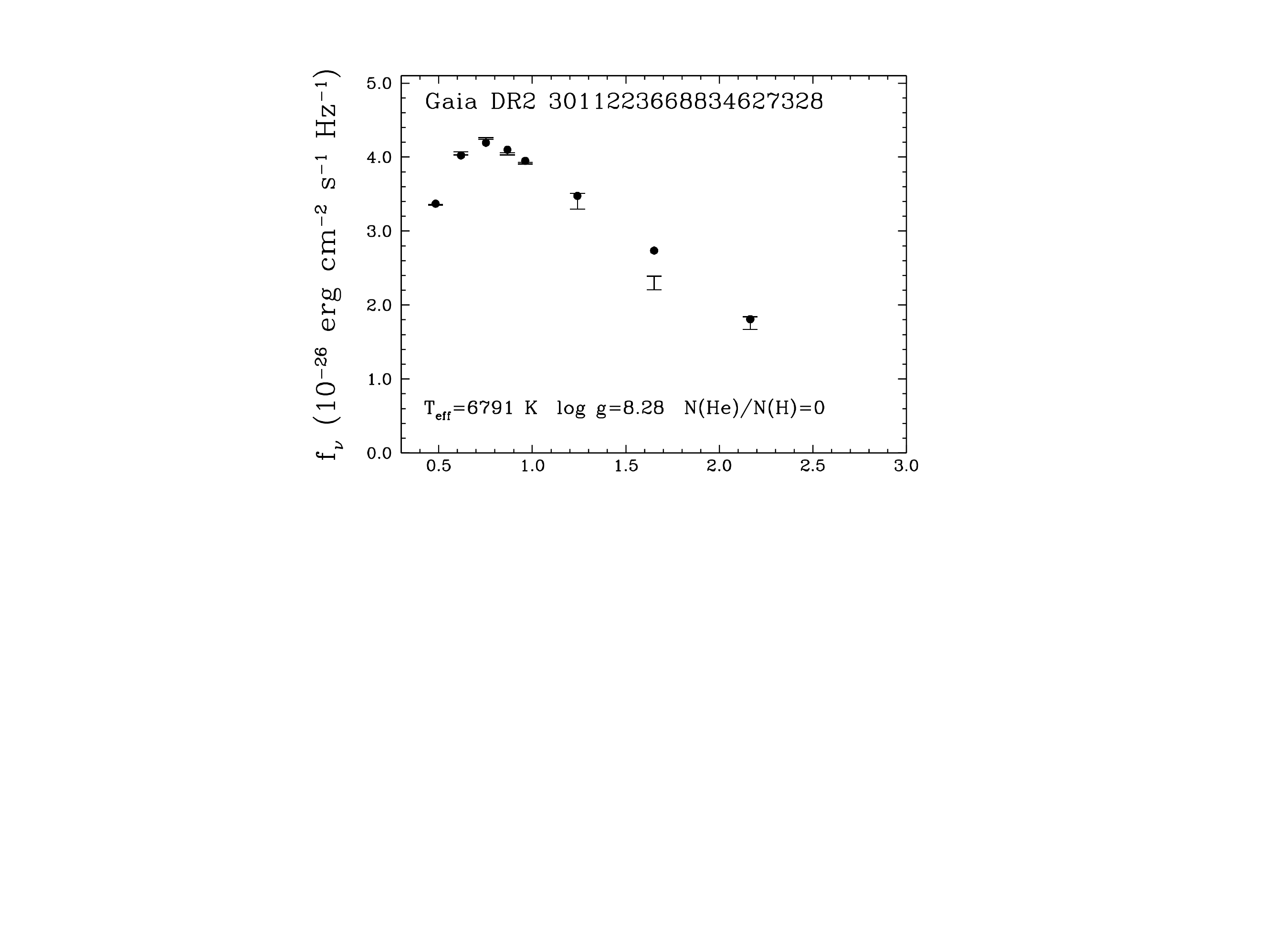}
\includegraphics[width=0.23\columnwidth,angle=270,bb = 10 145 260 490]{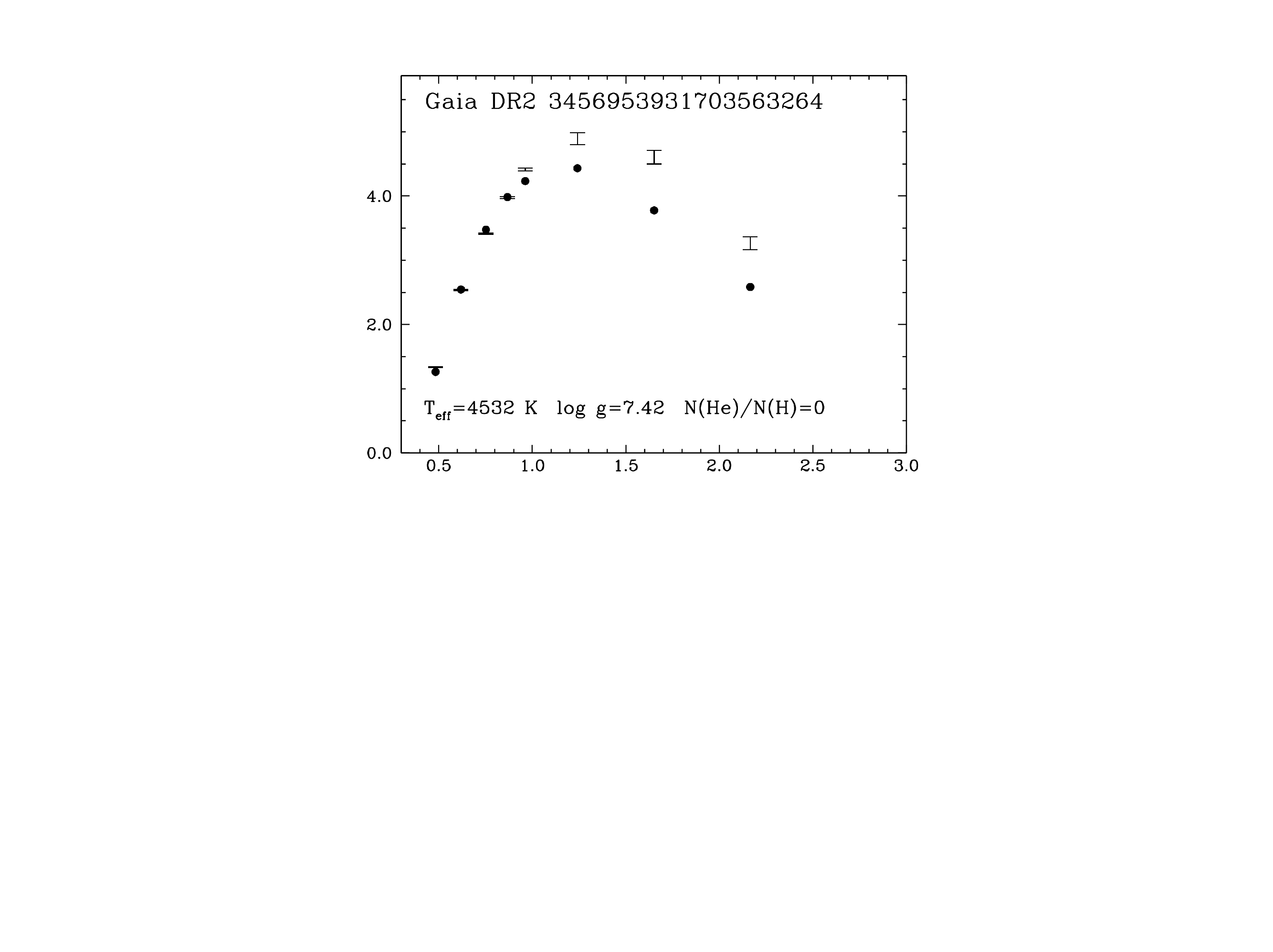}
\newline
\includegraphics[width=0.23\columnwidth,angle=270,bb = 10 145 260 490]{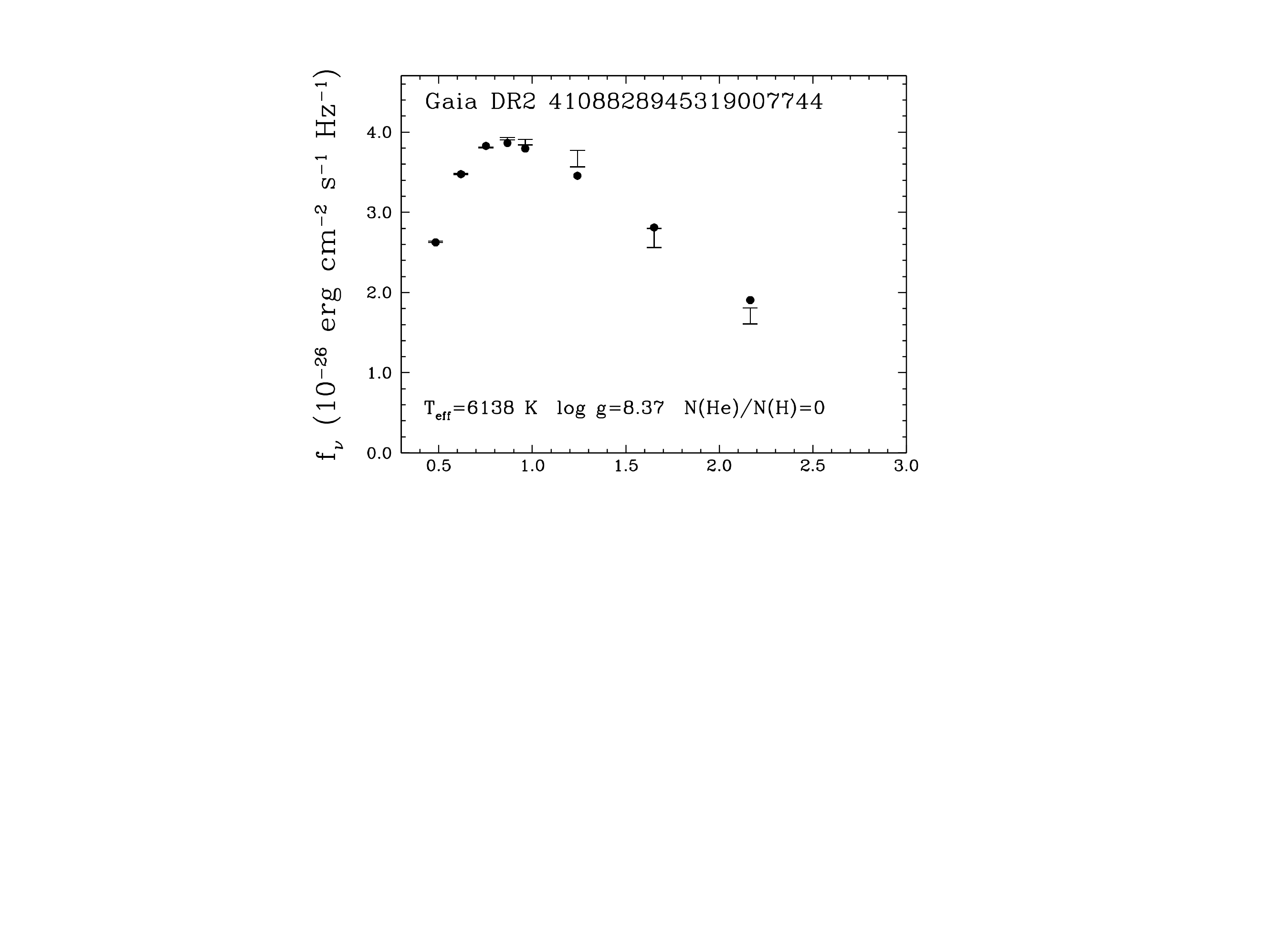}
\includegraphics[width=0.23\columnwidth,angle=270,bb = 10 145 260 490]{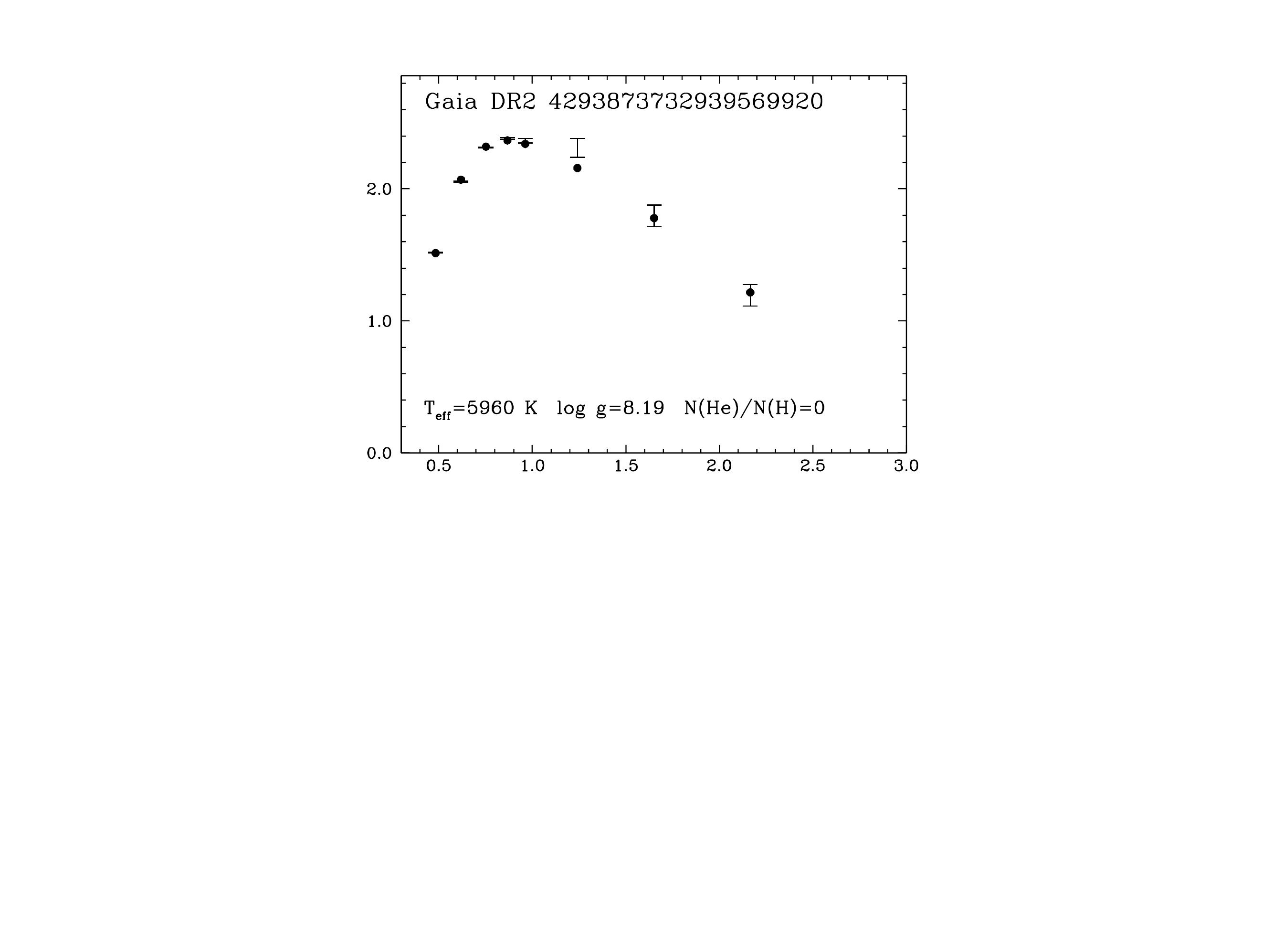}
\newline
\includegraphics[width=0.263\columnwidth,angle=270,bb = 10 145 295 490]{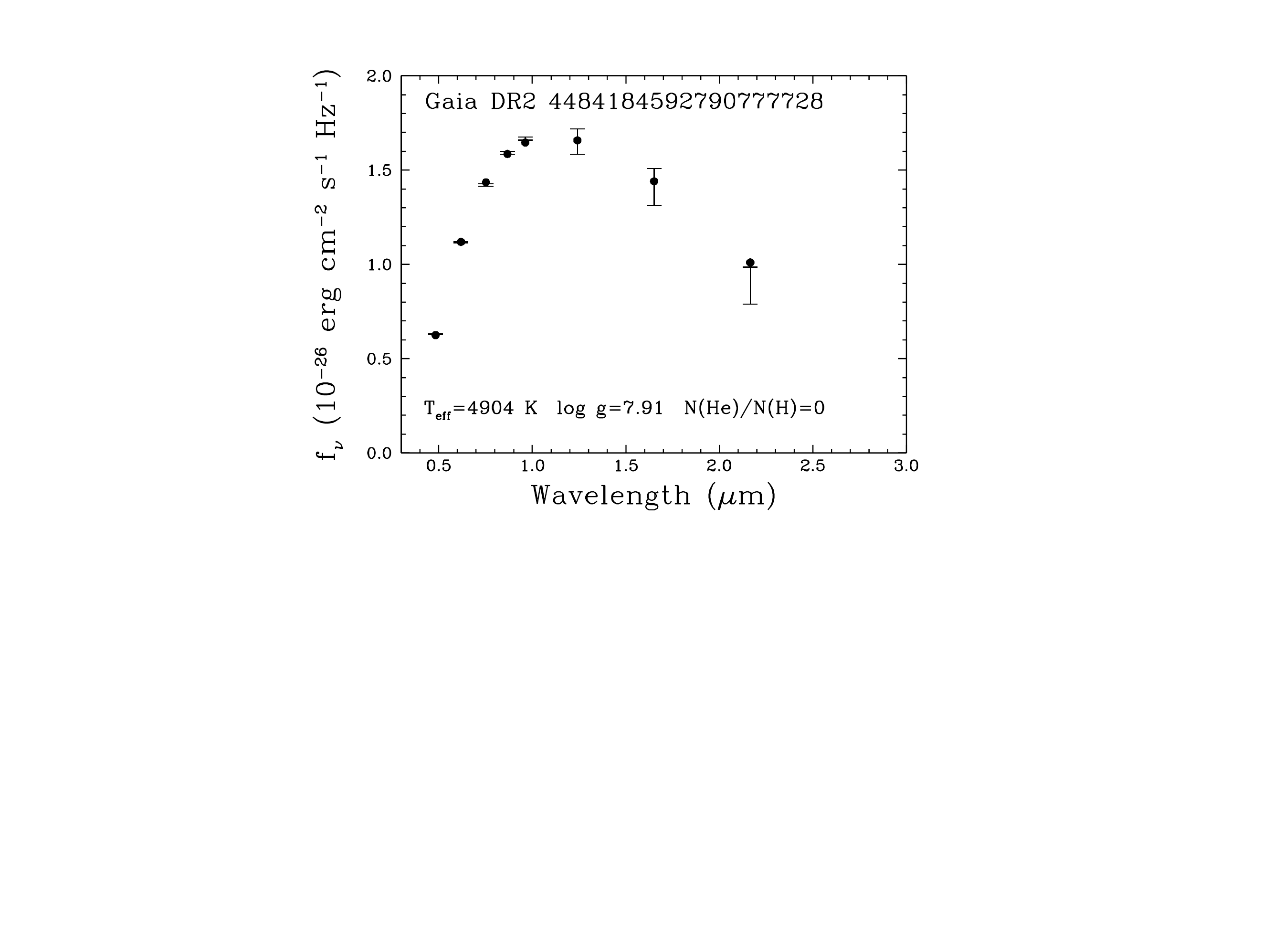}
\includegraphics[width=0.263\columnwidth,angle=270,bb = 10 145 295 490]{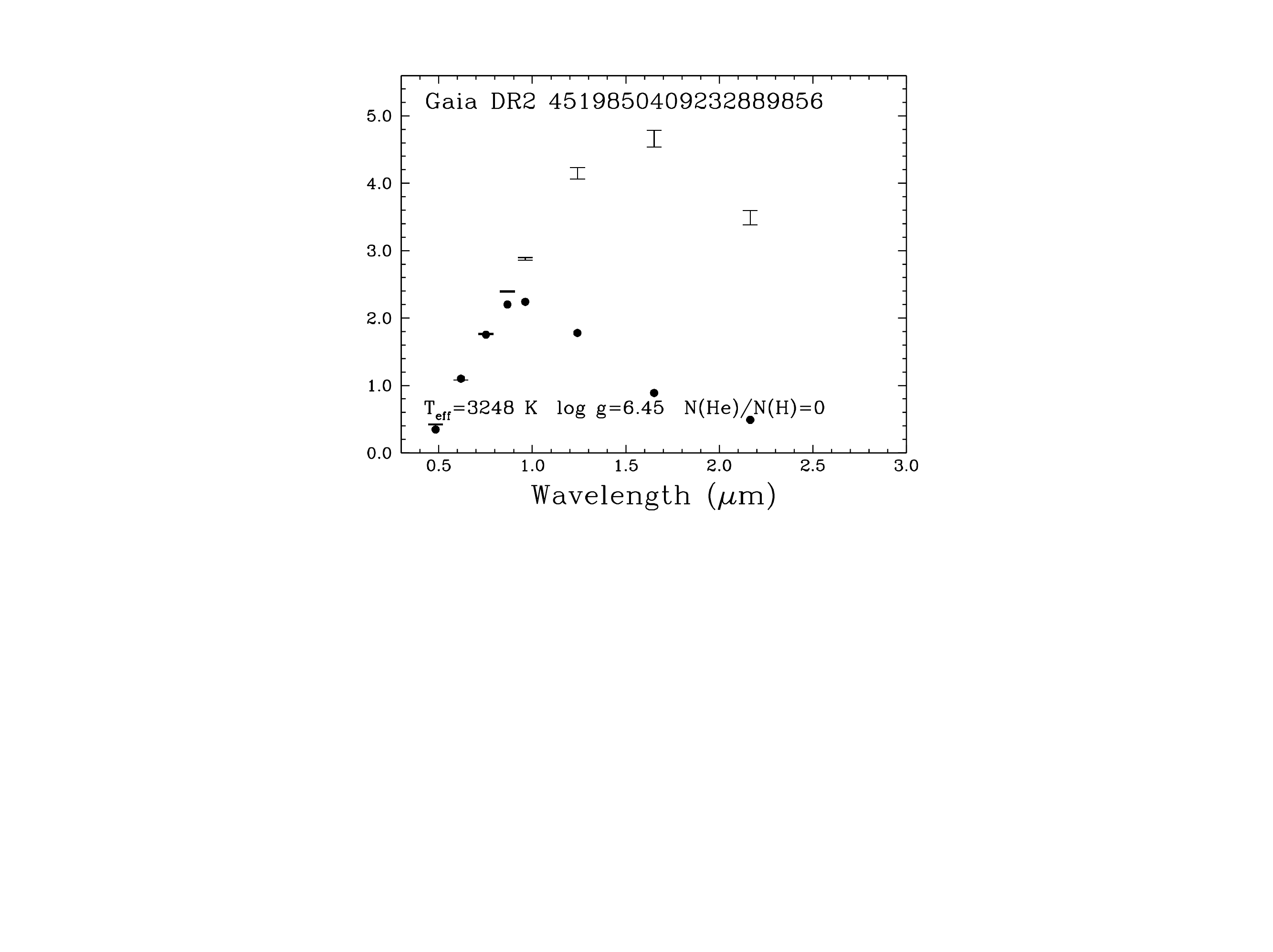}
\newline
\caption{Fits to Pan-STARRS and 2MASS photometry (error bars) of white dwarf candidates using \gaia\ parallaxes and assuming pure-H model atmospheres (solid points). \gaia\ DR2 1474900771698364672, 3456953931703563264, and 4519850409232889856 are unlikely to be white dwarfs (see Table~\ref{tab:rejected}).\label{fitsA1}}
\end{figure*}

\begin{figure*}
\centering
\includegraphics[width=0.23\columnwidth,angle=270,bb = 10 145 260 490]{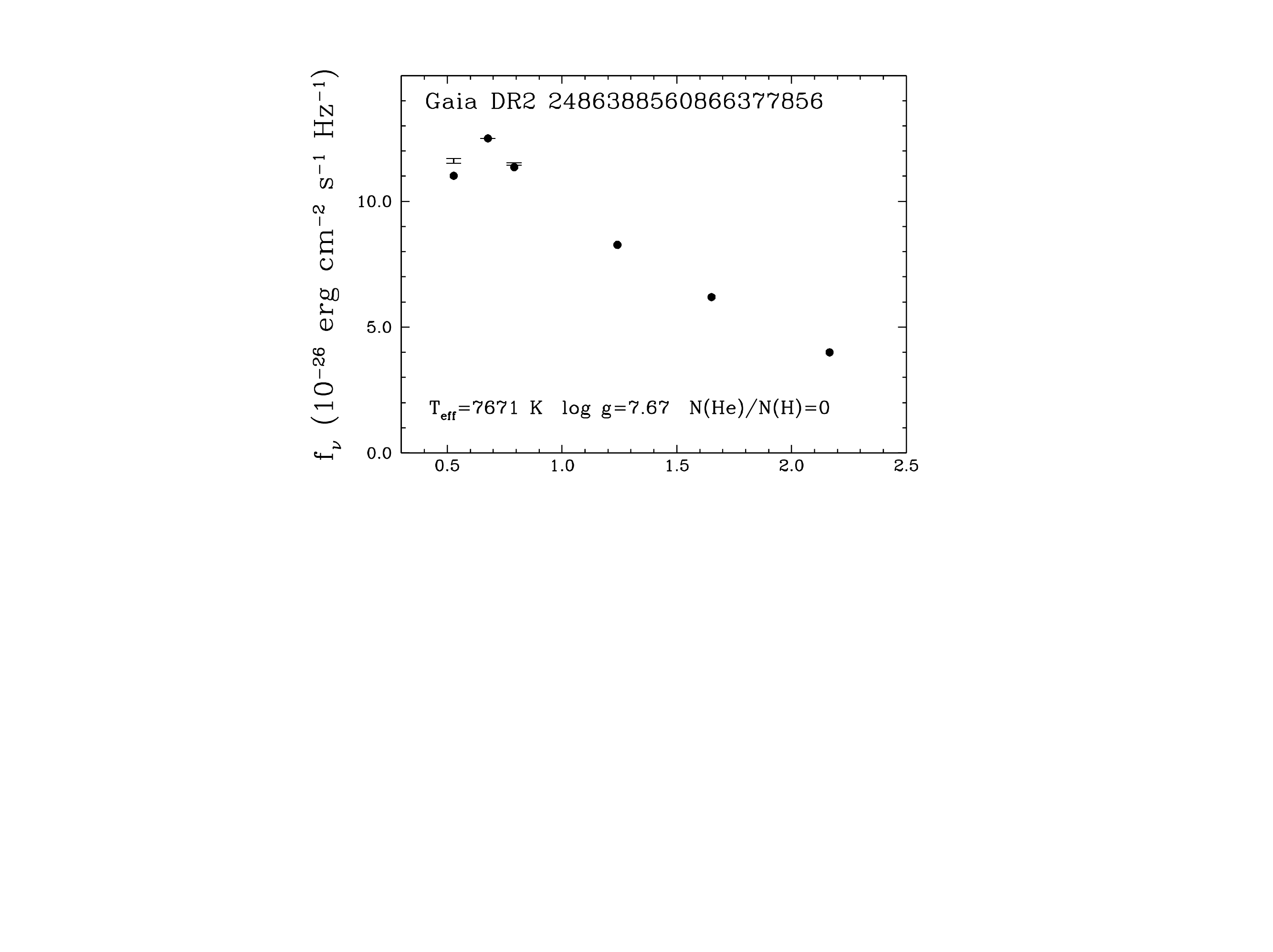}
\includegraphics[width=0.23\columnwidth,angle=270,bb = 10 145 260 490]{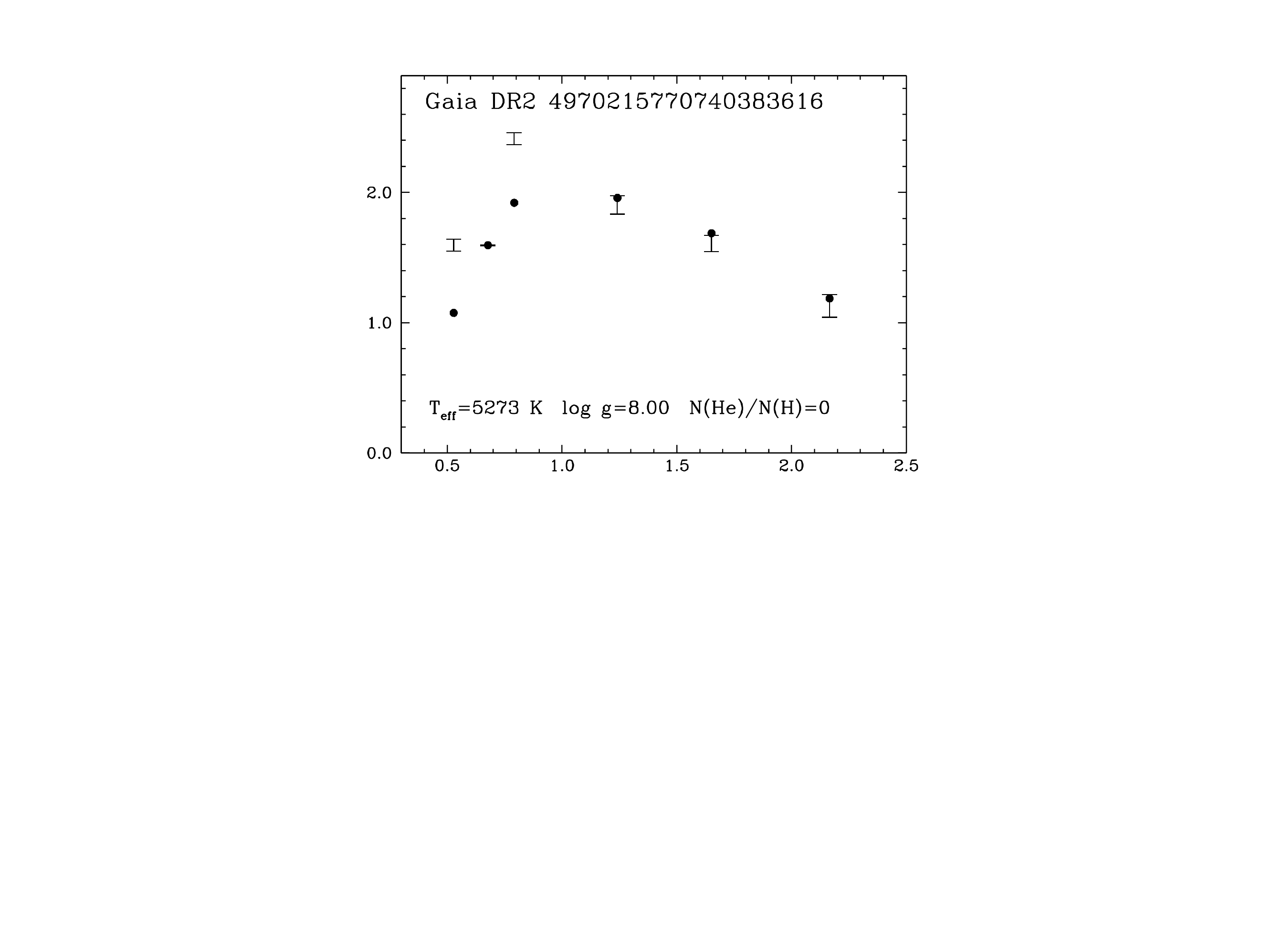}
\newline
\includegraphics[width=0.23\columnwidth,angle=270,bb = 10 145 260 490]{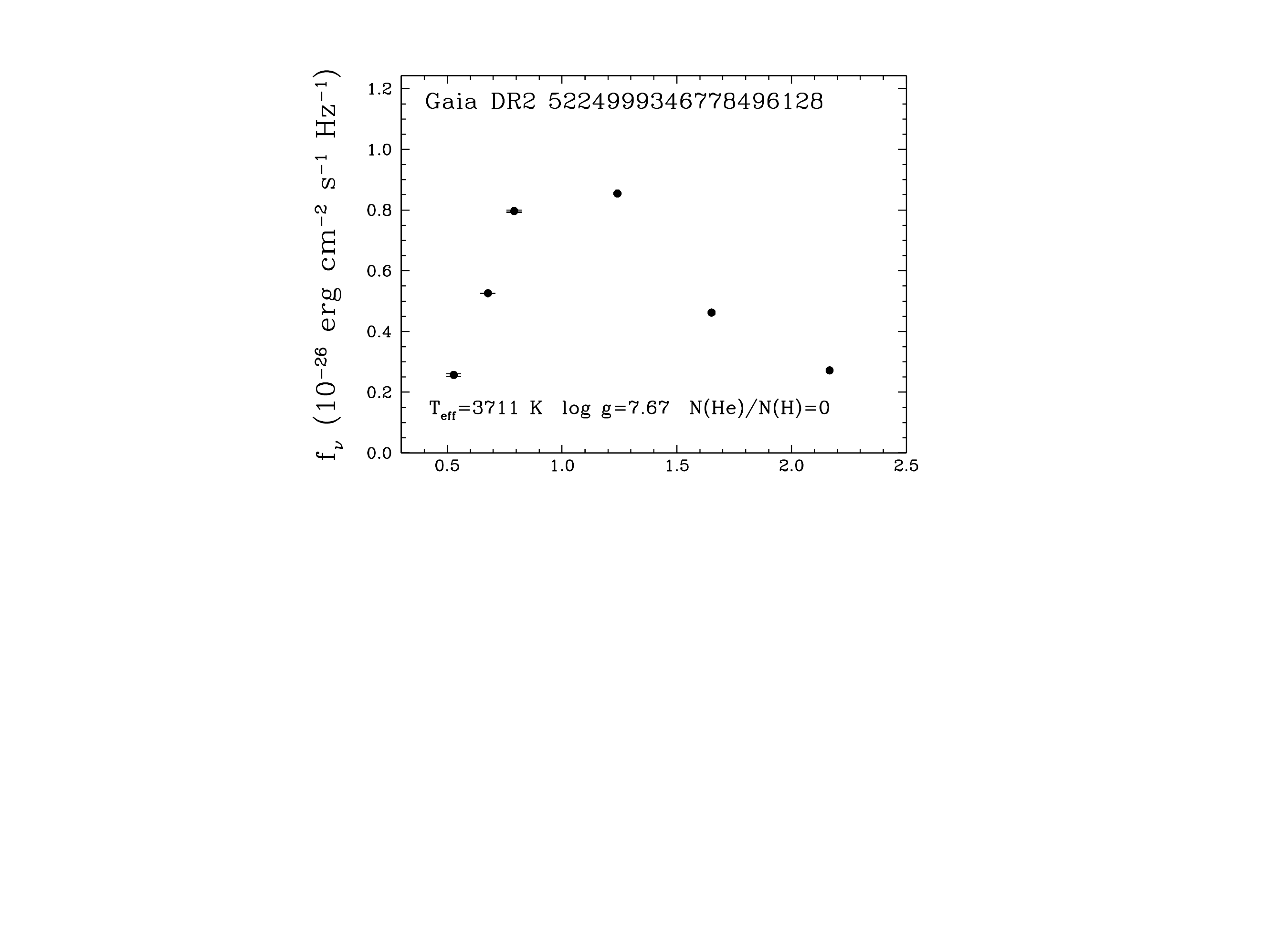}
\includegraphics[width=0.23\columnwidth,angle=270,bb = 10 145 260 490]{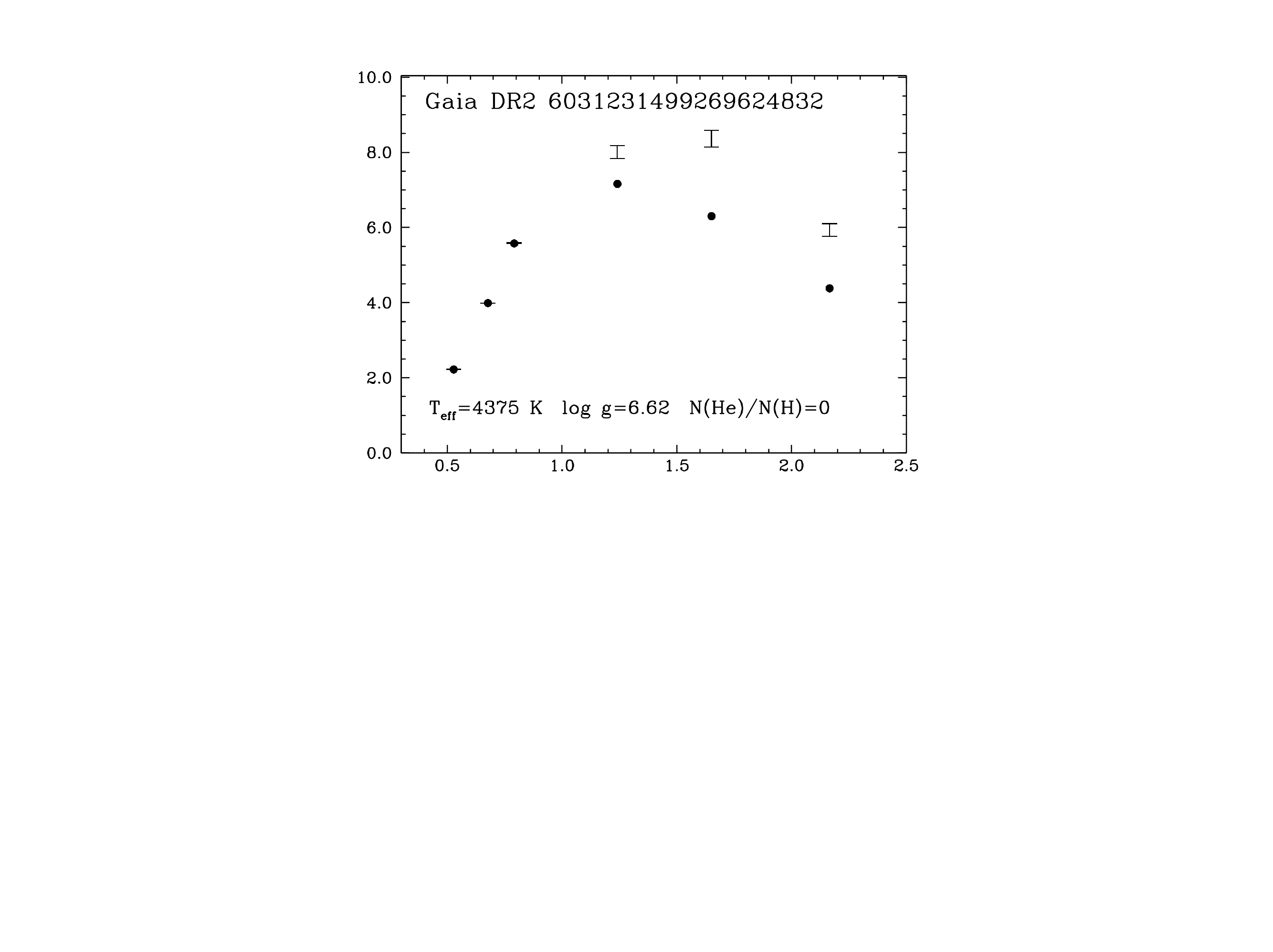}
\newline
\includegraphics[width=0.23\columnwidth,angle=270,bb = 10 145 260 490]{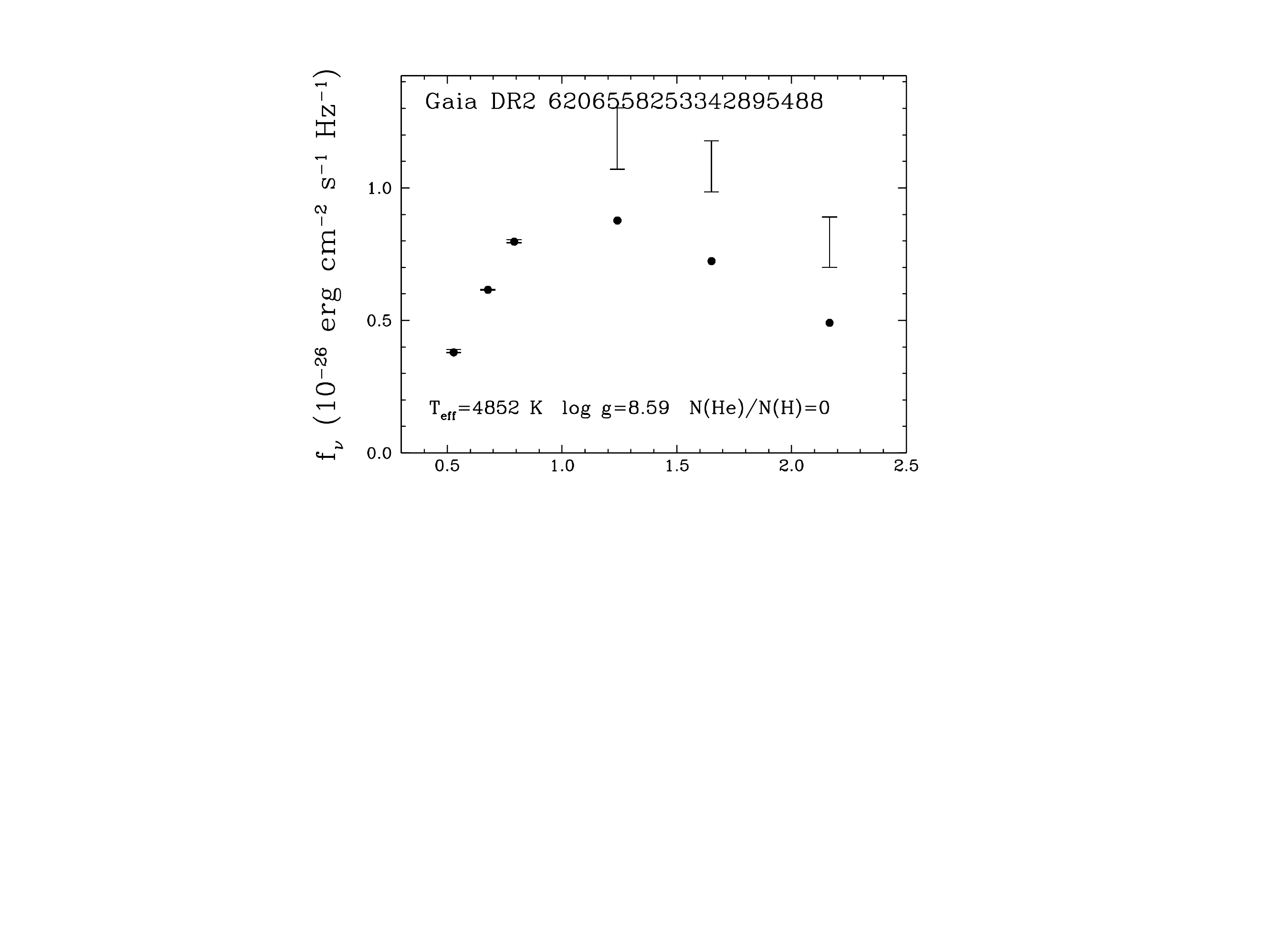}
\includegraphics[width=0.23\columnwidth,angle=270,bb = 10 145 260 490]{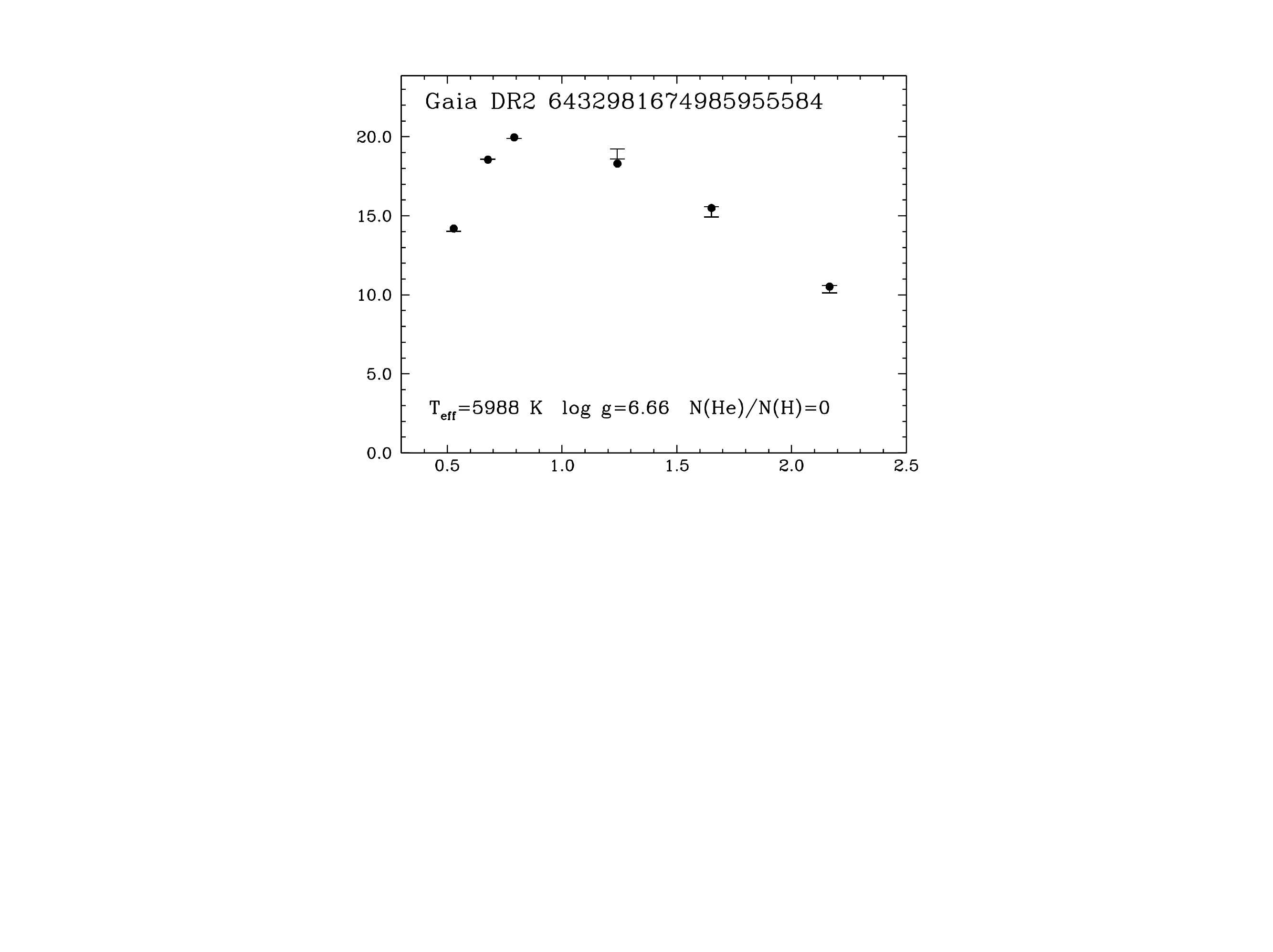}
\newline
\includegraphics[width=0.277\columnwidth,angle=270,bb = 10 145 310 490]{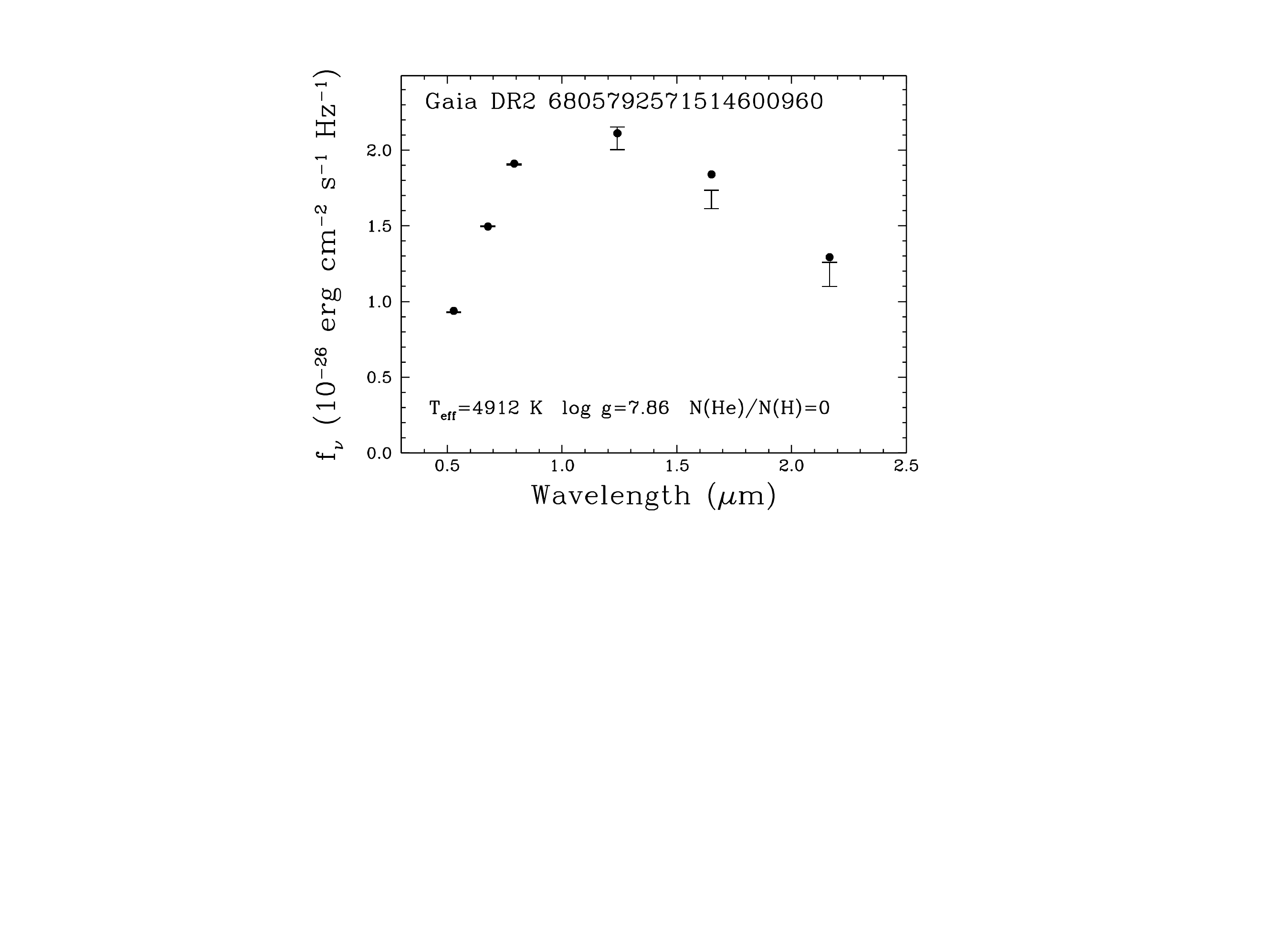}
\newline
\caption{Fit to \gaia\ and 2MASS photometry (error bars) for white dwarf candidates using \gaia\ parallaxes and assuming pure-H model atmospheres (solid points). For two sources, \gaia\ DR2 2486388560866377856 and 5224999346778496128, no 2MASS counterpart was found and we fit \gaia\ data only. 6031231499269624832, 6206558253342895488, and 6432981674985955584 are unlikely to be white dwarfs (see Table~\ref{tab:rejected}).\label{fitsA2}}
\end{figure*}

We display fits of 15 \gaia\ sources with $\textsc{astrometric\_excess\_noise$<$1}$\,mas that were selected by the colour cuts of Equations\,(\ref{Eq1}--\ref{Eq5}) but not matched with known white dwarfs. In Fig.~\ref{fitsA1} we show the objects that we found in both Pan-STARRS and 2MASS. In Fig.~\ref{fitsA2} we display the remaining sources that are not in Pan-STARRS and hence we use \gaia\ and 2MASS photometry. In all cases we employ the \gaia\ parallax to constrain the surface gravity.  We conclude that 9 objects are likely white dwarfs, which are catalogued in Table~\ref{tab:wd20_gaia} and discussed in Section~\ref{sec:newWDs}. The remaining 6 objects identified in Table~\ref{tab:rejected} are more likely to be main-sequence stars with erroneous \gaia\ astrometry and/or colours (see Section~\ref{sec:newWDs}).

\label{lastpage}
\end{document}